\documentclass[pre,twocolumn]{revtex4-2}

\usepackage{amsmath,amsfonts,amssymb}
\usepackage{amsthm}
\usepackage{bm}
\usepackage{mathrsfs}
\usepackage{graphicx}
\usepackage{hyperref}
\usepackage{braket}
\usepackage{physics}
\usepackage{mathtools}
\usepackage[whole]{bxcjkjatype}
\usepackage[normalem]{ulem}
\usepackage[usenames,dvipsnames]{xcolor}
\usepackage{soul}

\begin{document}

\title{Unified quantification of entanglement and nonstabilizerness \\in information scrambling and their trade-off relation} \author{Mao Kaneyasu}
\email{kaneyasu@biom.t.u-tokyo.ac.jp}
\affiliation{Department of Information and Communication Engineering, 
Graduate School of Information Science and Technology, The University of Tokyo,
Tokyo 113-8656, Japan}

\author{Yoshihiko Hasegawa}
\email{hasegawa@biom.t.u-tokyo.ac.jp}
\affiliation{Department of Information and Communication Engineering, 
Graduate School of Information Science and Technology, The University of Tokyo,
Tokyo 113-8656, Japan}
\date{\today}

\begin{abstract}

Entanglement and nonstabilizerness are among the most fundamental properties unique to quantum systems.
Each quantity captures a different aspect of non-classical behavior, and each can be regarded as a resource within its own operational setting.
However, the interrelation between them has not yet been fully clarified, and whether a more fundamental measure exists remains an open question.
Addressing these issues is essential for deepening our understanding of quantumness.
In this study, we establish a unified resource theory of information scrambling, consisting of two types: entanglement scrambling and nonstabilizerness scrambling.
We introduce a measure that jointly characterizes both types of scrambling.
This unified approach reveals a rigorous trade-off relation between entanglement and nonstabilizerness scrambling, as the exact maximum value of the proposed measure can be derived analytically.
Furthermore, we quantify the scrambling capability of unitary transformations in terms of their ability to amplify this measure.
We also discuss the relevance of our framework to the Pauli propagation method as a practical application.
Our work provides insights into the connection between entanglement and nonstabilizerness that extend beyond the context of information scrambling.

\end{abstract}

\maketitle

\section{Introduction}

Quantum systems exhibit a variety of non-classical and counterintuitive behaviors.
Understanding the origins of such quantumness remains one of the most fundamental challenges in quantum mechanics.
Quantum resource theories~\cite{horodecki2009quantum,bennett1996concentrating,vedral1998entanglement,lami2023no,baumgratz2014quantifying,winter2016operational,marvian2016quantum,napoli2016robustness,saxena2020dynamical,shiraishi2024arbitrary,theurer2017resource,guff2021resource,veitch2014resource,howard2017application,leone2022stabilizer,haug2023stabilizer,haug2023quantifying,leone2024stabilizer,zhang2024no,yunger2022resource,lostaglio2019introductory,brandao2013resource,liu2019one,regula2022probabilistic,zanoni2025choi} provide a powerful framework to capture and formalize the essential features of quantum systems (see Ref.~\cite{chitambar2019quantum} for a review).
In a typical framework, a fixed set of operations, known as \textit{free operations}, is specified in advance, and any property that cannot be generated using only these operations is regarded as a resource.
Constructing a resource theory for a given quantum property enables both its characterization and quantitative analysis.

Entanglement and nonstabilizerness, also known as magic, are key characteristics unique to quantum systems.
Each of these properties embodies quantumness, but they characterize different aspects, and their relationship remains unclear in several respects.
For instance, the GHZ state is highly entangled and exhibits strong quantum behavior.
However, it provides no quantum advantage from the perspective of stabilizer computation, as it can be generated solely by Clifford unitaries.
Conversely, the concatenation of T states exhibits large nonstabilizerness and enables quantum computational supremacy, even though it contains no entanglement.
From these observations, one natural question arises: \textit{is there any other crucial quantity that represents quantumness?}
Recent studies have explored the overlap between entanglement and nonstabilizerness~\cite{gu2025magic,dowling2025bridging,tirrito2024quantifying,tarabunga2025magic,szombathy2025asymptotically}.
Importantly, neither resource on its own can fully account for the source of quantum advantage in certain contexts, such as stabilizer tensor networks~\cite{masot2024stabilizer}.
The existence of a unified resource measure that incorporates both resources has been suggested, although its explicit form has not yet been specified~\cite{masot2024stabilizer,mittal2026quantum}.

\begin{figure}
  \includegraphics[width=8.0cm]{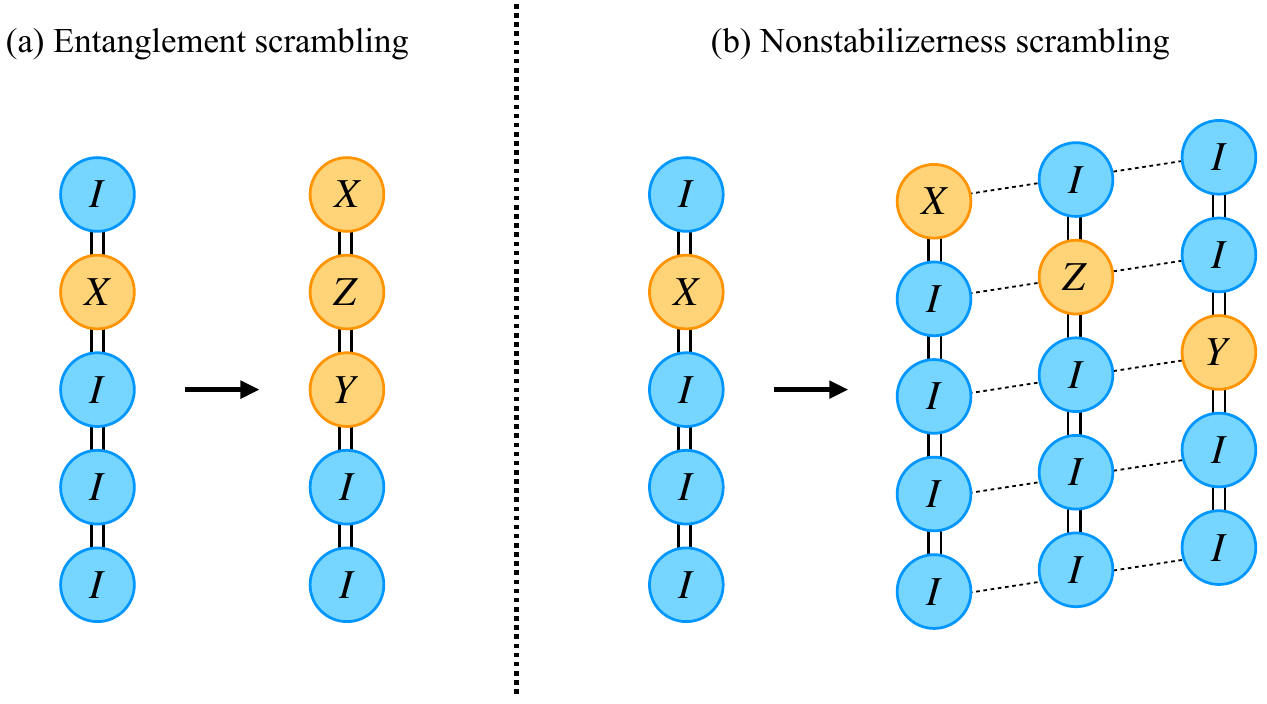}
  \caption{
  Schematic diagrams illustrating two types of information scrambling. (a) Entanglement scrambling corresponds to transformations that increase the number of non-identity operators in Pauli strings. (b) Nonstabilizerness scrambling corresponds to transformations that increase the number of terms required in the Pauli basis expansion.
  \label{fig:scrambling}
  }
\end{figure}

In this study, we establish a unified resource-theoretic framework of information scrambling and derive the rigorous trade-off relation between entanglement and nonstabilizerness within this framework.
Information scrambling refers to the phenomenon in which information initially localized in a small subsystem becomes distributed throughout the entire system~\cite{swingle2016measuring,landsman2019verified,bertini2020scrambling,mi2021information,zhou2017operator,knap2018entanglement,ahmadi2024quantifying,choi2020quantum,shen2020information,garcia2022quantifying,wu2021scrambling,mohseni2024deep,oliviero2024unscrambling}.
In the resource-theoretic context, scrambling is categorized into two distinct types based on their underlying mechanisms~\cite{garcia2023resource,bu2024complexity}: entanglement scrambling and nonstabilizerness scrambling (see FIG.~\ref{fig:scrambling}).
The latter was originally termed ``magic scrambling" in Ref.~\cite{garcia2023resource}.
Each type can be regarded as a resource under its own set of free operations.
However, treating them separately fails to reveal how they may influence or complement one another.
To clarify their interrelationship, we introduce a measure, termed \textit{operator complexity}, which captures both types of scrambling in a unified manner.
In particular, we emphasize that entanglement scrambling and nonstabilizerness scrambling cannot be simultaneously maximized.
As our most important result, we analytically derive the exact maximum value of the operator complexity, thereby revealing a fundamental trade-off inherent to information scrambling.
This discovery is expected to advance our understanding of how entanglement and nonstabilizerness are connected beyond the scope of information scrambling.
In addition, we propose a unified scrambling measure as an indicator of the ability to amplify the operator complexity.
These quantities satisfy several essential properties of resource monotones, ensuring the consistency and soundness of our framework.
Furthermore, to demonstrate the practical relevance of our unified framework, we review the Pauli propagation method \cite{rall2019simulation,rudolph2025pauli,angrisani2025classically,shao2024simulating,fontana2025classical}.
In this numerical method, the two types of scrambling are respectively related to the simulation complexity.
Thus, the proposed measures provide indicators of the computational cost in this setting.
Proofs for most propositions are provided in the Appendix.

\section{Scrambling of operators}
Information scrambling can be defined as the ability of unitary transformations to amplify the complexity of operators.
Entanglement scrambling and nonstabilizerness scrambling, which are different types of scrambling arising from distinct mechanisms, can be quantified using different measures of the complexity of operators.
In this section, we recall two measures of operators that correspond to each scrambling in the context of resource theory.
Furthermore, we introduce a unified measure that takes into account both scrambling and derive its rigorous maximum value.
This clarifies the fundamental trade-off between the two types of scrambling.

We consider an $n$-qubit system in the $2^n$-dimensional Hilbert space $\mathcal{H}$.
Let $O$ be a normalized operator acting on $\mathcal{H}$, which can be expanded in the Pauli basis as $O = \sum_i c_iP_i$.
Here, $\{P_i\}_i$ denotes the subset of the Pauli group on $\mathcal{H}$ consisting of elements with coefficients equal to $+1$.
The coefficients satisfy the condition $\sum_i|c_i|^2=1$, ensuring that the norm of $O$ is normalized to unity.

Firstly, we focus on entanglement scrambling.
As illustrated in FIG.~\ref{fig:scrambling}, entanglement scrambling corresponds to transformations that increase the number of non-identity operators in the Pauli strings.
To quantify this spreading, we define the average Pauli weight \cite{garcia2023resource,bu2024complexity,roberts2018operator,qi2019measuring} of the normalized operator $O$ as
\begin{equation}
    W(O) \coloneqq \sum_i |c_i|^2W(P_i).
\end{equation}
For any Pauli operator $P$, $W(P)$ denotes the number of non-identity operators composing the Pauli string.
For any normalized operator $O$, the average Pauli weight has the following basic properties:
\begin{enumerate}
    \item Faithfulness: $W(O)\geq 0$, and $W(O)=0$ if and only if $O$ is the identity operator.
    \item[$\tilde{1}$.] Faithfulness (traceless case): $W(O)\geq 1$ for any traceless $O$, and $W(O)=1$ if and only if $O$ can be expanded using only weight-1 Pauli operators.
    \item Invariance: $W(U^\dagger OU) = W(O)$ for any non-entangling unitary $U$.
    \item Additivity: $W(O_1\otimes O_2) = W(O_1) + W(O_2)$ for any normalized operators $O_1$ and $O_2$.
\end{enumerate}
Here, a non-entangling unitary refers to a unitary that can be generated from single-qubit unitaries and swap operations.
Thus, $W(O)$ can be regarded as a resource in the same setting as the state-based entanglement, where the set of non-entangling unitaries constitutes the free operations.
This observation suggests a connection between the scrambling characterized by $W(O)$ and state-based entanglement, which motivates the term \textit{entanglement scrambling}.
More broadly, resource-theoretic frameworks can capture the essence of quantum properties.
Besides, we note that $W(O)$ is defined without reference to any particular bipartition, whereas state-based entanglement measures, such as entanglement entropy, generally depend on the choice of bipartition.
Since scrambling is not inherently associated with a specific bipartition, this feature makes $W(O)$ well suited for quantifying the scrambling capability of unitary transformations.

Next, we consider the quantification of nonstabilizerness scrambling, which corresponds to an increase in the number of terms in the Pauli basis expansion of the operator (see FIG.~\ref{fig:scrambling}).
We define a measure of the complexity of the Pauli basis expansion, called the quantum Fourier entropy \cite{bu2024complexity}, as
\begin{equation}
    S_a(O) \coloneqq -\sum_i |c_i|^2\log_a |c_i|^2.
\end{equation}
The advantages of this operator-based measure are discussed in Ref.~\cite{dowling2025magic}.
Here, we take the base of the logarithm as a parameter $a$.
If $a$ is omitted, we set $a=4$ so that the maximum value of the quantum Fourier entropy becomes $n$.
This choice aligns the ranges of $W(O)$ and $S(O)$, enabling us to treat the two types of scrambling on equal footing.
For any normalized operator $O$, the quantum Fourier entropy satisfies the following properties:
\begin{enumerate}
    \item Faithfulness: $S_a(O)\geq 0$, and $S_a(O)=0$ if and only if $O$ is a Pauli operator.
    \item Invariance: $S_a(U^\dagger OU) = S_a(O)$ for any Clifford unitary $U$.
    \item Additivity: $S_a(O_1\otimes O_2) = S_a(O_1) + S_a(O_2)$ for any normalized operators $O_1$ and $O_2$.
\end{enumerate}
Thus, $S_a(O)$ can be regarded as a resource when Clifford unitaries are taken as free operations.
Accordingly, this type of scrambling is referred to as \textit{nonstabilizerness scrambling}.

Here, we introduce a resource-theoretic measure of the complexity of operators.
Since this function captures both types of scrambling, it enables a unified treatment of information scrambling:
\begin{equation}
\label{def_r}
    R_a(O) \coloneqq W(O) + S_a(O) = \sum_i |c_i|^2\log_a\frac{a^{W(P_i)}}{|c_i|^2}.
\end{equation}
We refer to this measure as \textit{operator complexity}.
The main contribution of this study is the introduction and analysis of the operator complexity.
$R_a(O)$ quantifies the extent to which a normalized operator simultaneously exhibits entanglement and nonstabilizerness scrambling.
Since $W(O)$ and $S(O)$ share the same range of values, when the parameter $a$ is set to 4, Eq. \eqref{def_r} treats both types of scrambling equally.
For any normalized operator $O$, the operator complexity satisfies the following properties:
\begin{enumerate}
    \item Faithfulness: $R_a(O)\geq 0$, and $R_a(O)=0$ if and only if $O$ is the identity operator.
    \item[$\tilde{1}$.] Faithfulness (traceless case): $R_a(O)\geq 1$ for any traceless $O$, and $R_a(O)=1$ if and only if $O$ is a weight-1 Pauli operator.
    \item Invariance: $R_a(U^\dagger OU) = R_a(O)$ for any non-scrambling unitary $U$.
    \item Additivity: $R_a(O_1\otimes O_2) = R_a(O_1) + R_a(O_2)$ for any normalized operators $O_1$ and $O_2$.
\end{enumerate}
Here, non-scrambling unitaries are defined as the intersection of non-entangling unitaries and Clifford unitaries; that is, any such unitary can be generated from single-qubit Clifford unitaries and swap operations.
These properties ensure that $R_a(O)$ behaves as a valid resource monotone.
Thus, verifying such properties is essential in the context of resource theories.
They follow directly from the properties of the average Pauli weight and the quantum Fourier entropy, and are built into our construction.
In particular, the additivity is a key property that makes $R_a(O)$ more tractable, as it inherits this property from both $W(O)$ and $S_a(O)$ by design.
This is precisely the reason why we adopt the present definition of $R_a(O)$.

\begin{figure}
  \includegraphics[width=8.5cm]{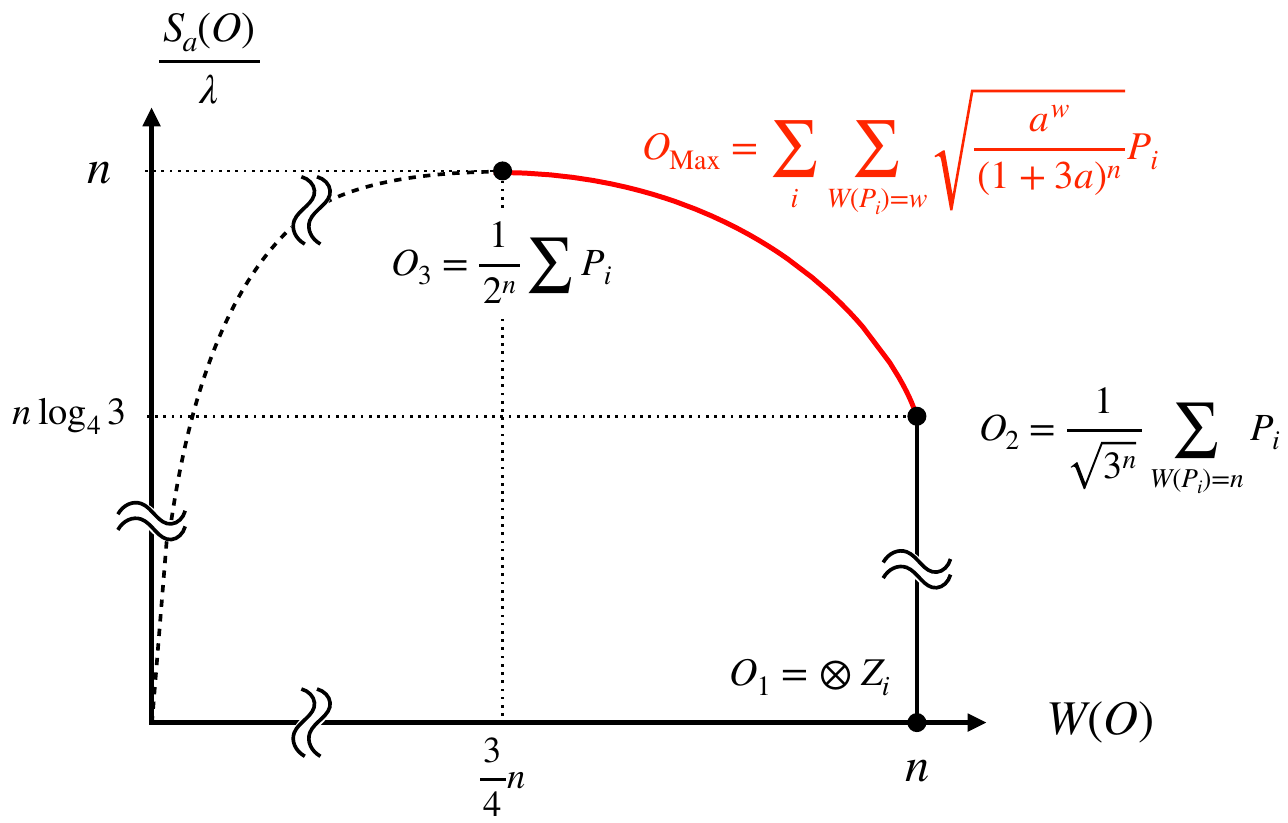}
  \caption{
  Overview of the $W$-$S$ plane for arbitrary linear operators. A key observation is the trade-off relation between the two types of scrambling, illustrated by the solid red curve. When $W(O)$ reaches its maximum, $S_a(O)$ cannot, and vice versa: when $S_a(O)$ is maximized, $W(O)$ takes a specific, non-maximal value. The dashed curve provides a rough sketch of the general trend (possibly imprecise).
  \label{fig:mapping}
  }
\end{figure}

$W(O)$ and $S_a(O)$ cannot be simultaneously maximized; that is, there exists a trade-off between entanglement scrambling and nonstabilizerness scrambling.
Here, we derive the exact maximum value of the operator complexity.
This result reveals a fundamental limitation inherent in information scrambling.

Since the value of the operator complexity is independent of the complex phases of the coefficients, we represent the normalized operator $O$ as $O = \sum_i \sqrt{p_i}P_i$.
Under the constraints $\sum_ip_i = 1$ and $p_i\geq 0$, we solve the following optimization problem:
\begin{equation}
    \mathrm{maximize} \;\; R_a(O) = \sum_i p_i \left\{ W(P_i) -\log_a p_i \right\}.
\end{equation}
We provide the solution to this problem in the Appendix.
As a result, the closed-form expression of the operator that maximizes $R_a(O)$ is derived as
\begin{equation}
    O_\mathrm{Max} = \sum_{w=0}^n \sum_{W(P_i)=w} e^{i\theta_i}\sqrt{\frac{a^w}{(1+3a)^n}}P_i,
\end{equation}
where $\{\theta_i\}_i$ are arbitrary phases.
This operator exhibits optimal spreading from both scrambling perspectives.
Furthermore, the values of each measure obtained from $O_\mathrm{Max}$ can be computed as follows:
\begin{equation}
\label{WOmax}
    W(O_\mathrm{Max}) = \frac{3a}{1+3a}n,
\end{equation}
\begin{equation}
\label{SOmax}
    S_a(O_\mathrm{Max}) = n\log_a(1+3a) - \frac{3a}{1+3a}n,
\end{equation}
\begin{equation}
    R_a(O_\mathrm{Max}) = n\log_a(1+3a).
\end{equation}
See the Appendix for the details of the calculation.
In particular, when the parameter $a$ is set to 4, we obtain
\begin{equation}
\label{ROmax}
    R(O) \leq R(O_{\mathrm{Max}}) \simeq 1.85n < 2n.
\end{equation}
This is a key finding that stems from the operator complexity.
Since the maximum values of $W(O)$ and $S(O)$ are both $n$, the operator complexity is clearly bounded above by $2n$.
However, Eq. \eqref{ROmax} reveals the upper bound strictly smaller than $2n$.
This bound represents the exact maximum of the operator complexity, reflecting a fundamental constraint inherent in scrambling dynamics.
Similarly, for general values of the parameter $a$, we obtain the following inequality:
\begin{equation}
\label{RaOmax}
    R_a(O) \leq R_a(O_\mathrm{Max}) = n\log_a(1+3a) < (1+\lambda)n,
\end{equation}
where $\lambda = \log_a4$. The quantity $(1+\lambda)n$ is a trivial upper bound, since $R_a(O) = W(O)+\lambda S(O)$. Eq.~\eqref{RaOmax} also captures the intrinsic limitation between the two types of scrambling.
Additionally, we point out that a tensor product of single-qubit maximally scrambled operators $O^1_\mathrm{Max}$ yields a maximally scrambled operator on a multi-qubit system:
\begin{equation}
    \bigotimes_{i=1}^n \;O^{(1)}_{\mathrm{Max},i} = O_\mathrm{Max}^{(n)}.
\end{equation}
This is a fundamental property inherited from the additivity of $R_a(O)$, which is built into our definition.

In FIG.~\ref{fig:mapping}, we map the several representative operators on the $W$-$S$ plane.
First, any weight-$n$ Pauli operator, such as $O_1 = Z_1\otimes\cdots\otimes Z_n$, gives the maximum value of $W(O)$, while $S_a(O)$ remains zero.
Linear combinations of weight-$n$ Pauli operators also attain the maximum value of $W(O)$, and although they yield nonzero values of $S_a(O)$, they do not maximize it.
For instance, the operator $O_2$ shown in FIG.~\ref{fig:mapping} maximizes $W(O)$ and achieves $S_a(O_2) = n\log_a3$, which is smaller than the maximum value of $S_a(O)$.
In contrast, the uniform linear combination of all Pauli basis elements with equal amplitude, $O_3 = 1/2^n\sum_iP_i$, maximizes $S_a(O)$, while its average Pauli weight is not maximized: $W(O_3)=3n/4$.
Most importantly, $O_\mathrm{Max}$ is optimal with respect to the operator complexity $R_a(O)$, even though it does not individually maximize $W(O)$ or $S_a(O)$.
The exact location of $O_\mathrm{Max}$ as a function of the parameter $a$ is also illustrated in FIG~\ref{fig:mapping}.
In the limit $a\rightarrow1$, $O_\mathrm{Max}$ converges to $O_3$, whereas in the limit $a\rightarrow\infty$, it converges to $O_2$.

\begin{widetext}

{
\tabcolsep = 20pt
\renewcommand{\arraystretch}{1.8}

\begin{table}[]
\centering
\label{table:view}
\caption{
Overview of the measures for each type of information scrambling. A parallel structure emerges based on the resource-theoretic framework, with each type of scrambling associated with its own set of free operations.
}
\vspace{0.1in}
\begin{tabular}{|c|c|c|} \hline
Operator & Unitary & Free unitary \\ \hline\hline
$W(O) = \sum_i|c_i|^2W(P_i)$&
$G_E(U)\coloneqq\max \left[ W(U^\dagger OU) - W(O) \right]$&
Non-entangling unitary \\ \hline
$S(O)=-\sum_i|c_i|^2\log|c_i|^2$&
$G_M(U)\coloneqq\max \left[ S(U^\dagger OU) - S(O) \right]$&
Clifford unitary \\ \hline
$R(O)=\sum_i|c_i|^2(W(P_i)-\log|c_i|^2)$&
$G(U)\coloneqq\max \left[ R(U^\dagger OU) - R(O) \right]$&
Non-scrambling unitary \\ \hline
\end{tabular}
\end{table}
}
    
\end{widetext}

\section{Scrambling of unitaries}
The operator-based measures discussed in the previous section can be used to quantify information scrambling of unitary transformations via their action on operators.
The average Pauli weight and the quantum Fourier entropy are associated with entanglement scrambling and nonstabilizerness scrambling, respectively.
Furthermore, the operator complexity provides a unified framework for quantifying information scrambling.

We define the following measures corresponding to each type of scrambling: the \textit{entanglement growth}, the \textit{nonstabilizerness growth}, and the \textit{complexity growth}, the latter being introduced in this work.
\begin{equation}
\label{def_eg}
    G_E(U) \coloneqq \max_{ \|O\|_2=1 } \left\{ W(U^\dagger OU) - W(O) \right\},
\end{equation}
\begin{equation}
\label{def_mg}
    G_M(U) \coloneqq \max_{\|O\|_2=1} \left\{ S(U^\dagger OU) - S(O) \right\},
\end{equation}
\begin{equation}
\label{def_cg}
    G(U) \coloneqq \max_{ \|O\|_2=1 } \left\{ R(U^\dagger OU) - R(O) \right\}.
\end{equation}
The definitions presented here are inspired by the ideas discussed in Refs.~\cite{garcia2023resource,bu2024complexity}, where the corresponding definitions include an absolute value.
The use of the absolute value imposes a symmetry such as $G(U)=G(U^\dagger)$, which is not necessarily problematic.
However, since it may hinder further extensions of our framework (e.g., to general TPCP channels), we slightly modify the definitions by removing the absolute value, with the expectation this leads to greater flexibility.
We also note that the essential properties as resource monotones remain unchanged regardless of which definition is adopted.
Here, we set the parameter $a$ to 4, but the definition extends naturally to general values.
These functions satisfy the following fundamental properties:
\begin{enumerate}
    \item Faithfulness: $G_{(E,M)}(U)\geq 0$, and $G_{(E,M)}(U)=0$ if and only if $U$ is a free unitary.
    \item Invariance: $G_{(E,M)}(V_1UV_2) = G_{(E,M)}(U)$ for any free unitaries $V_1, V_2$.
    \item Subadditivity: $G_{(E,M)}(U_1\otimes U_2) \leq G_{(E,M)}(U_1) + G_{(E,M)}(U_2)$ for any unitaries $U_1$ and $U_2$.
\end{enumerate}
Non-entangling unitaries, Clifford unitaries, and non-scrambling unitaries constitute the set of free unitaries for $G_E(U)$, $G_M(U)$, and $G(U)$, respectively.
Thus, these serve as valid measures from the perspective of resource theory.

Here, we provide an overview of the resource-theoretic framework for information scrambling (see TABLE 1.).
We present three distinct operator-based measures; $W(O)$, $S(O)$, and $R(O)$.
In particular, $R(O)$ captures both entanglement and nonstabilizerness scrambling.
Based on the ability of unitary transformations to amplify these scrambling-related features, we define corresponding measures of information scrambling.
These measures satisfy several fundamental properties, allowing them to be treated as resources.
For each type of scrambling, an appropriate set of free unitaries is identified.
Taken together, the framework exhibits a parallel structure across the different scrambling types.

\section{Application to Pauli propagation}
The proposed measure of unified scrambling can serve as a resource in the context of numerical simulations.
Here, we briefly review the Pauli propagation method \cite{rall2019simulation,rudolph2025pauli,angrisani2025classically,shao2024simulating,fontana2025classical} and discuss its relation to the measure of unified scrambling.

This is a classical method for simulating quantum circuit.
We consider a multi-qubit system initialized in a state $\rho_0$ and an arbitrary observable $O$.
Our goal is to compute the expectation value of $O$ after a unitary transformation $U$, given by $\mathrm{Tr}[\rho_0(U^\dagger OU)]$, where we work in the Heisenberg picture.
The observable $O$ can be expanded in the Pauli basis as $O = \sum_i c_iP_i$.
By tracking these coefficients and the corresponding Pauli operators, we can simulate the evolution of the observable.

In general, any unitary can be decomposed into a product of Clifford unitaries and Pauli rotations.
Since Clifford unitaries map Pauli operators to other Pauli operators, their action can be simulated by simply swapping the coefficients.
In contrast, Pauli rotations with rotation angle $\theta$ can induce branching as follows \cite{rudolph2025pauli}:
\begin{equation}
\label{branching}
e^{i\frac{\theta}{2} G} P e^{-i\frac{\theta}{2} G} =
    \begin{cases}
    P & \text{if $[P,G]=0$} \\
    \cos\theta P + \sin\theta P' & \text{else}
  \end{cases}
\end{equation}
where $P$ and $G$ are Pauli operators, and $P' = i[P,G]/2$ is another Pauli operator.
As this branching process is repeated, the simulation becomes increasingly complex.

We can estimate the complexity of the simulation by considering both types of scrambling.
First, the larger the number of terms in the Pauli expansion, the more complex the simulation becomes.
This can be quantified by the quantum Fourier entropy $S_a(O)$, which reflects the number of terms in the Pauli expansion.
In the Pauli propagation method, terms with small coefficients are typically truncated to reduce computational cost.
Thus, $S_a(O)$ provides a more informative indicator of the complexity than merely counting the number of terms.
Next, we note that the branching in Eq.~\eqref{branching} occurs only when two Pauli operators $P$ and $G$ anti-commute.
If $P$ is a tensor product of non-identity Pauli operators, it anti-commutes with a large number of local Pauli operators, which are typically used in practical quantum circuits.
In contrast, if $P$ mostly consists of identity operators, it commutes with most local Pauli operators.
Therefore, operators with many non-identity components tend to result in more complex simulations, which is captured by the average Pauli weight $W(O)$.
Hence, the operator complexity $R_a(O) = W(O)+S_a(O)$ can serve as a measure of the computational complexity of Pauli propagation.
Furthermore, we can interpret the scrambling measure for unitaries $G(U) = \max_O \{ R(U^\dagger OU) - R(O) \}$ as an indicator of the worst-case computational complexity for a given quantum circuit $U$.
It is expected that improving truncation strategies from the perspective of the unified scrambling measures will reduce computational cost of the Pauli propagation method.

\section{Conclusion}
In this study, we constructed a unified framework of resource theory for information scrambling.
In particular, we introduced the concept of operator complexity, which accounts both entanglement scrambling and nonstabilizerness scrambling.
As a main result, we derived the exact maximum value of this measure, revealing a trade-off relation inherent in information scrambling.
In addition, we quantified the scrambling capability intrinsic to unitary transformations as the amplification of such measures.
These measures exhibit a parallel structure and satisfy the properties required of resource monotones.
Furthermore, as a practical application of our framework, we examined the relation between our unified measure and the Pauli propagation method.

As for future directions, developing efficient methods to compute concrete values of scrambling measures will be an important task.
Furthermore, our framework can likely be extended to non-unitary channels, allowing the quantification of their scrambling capability.
We expect that this work will deepen the understanding of the fundamental nature of information scrambling.
Moreover, our analysis offers perspectives on the connection between entanglement and nonstabilizerness that go beyond the scope of scrambling itself.
We hope that this study will inspire further developments across a broad range of topics in quantum physics.

\begin{acknowledgments}

This work was supported by JSPS KAKENHI Grant Numbers JP24KJ0960, JP23K24915 and JP24K03008.

\end{acknowledgments}

\section*{Data Availability}
The code used to generate the trade-off curve presented in this work is publicly available \cite{availablity}.

\appendix
\setcounter{secnumdepth}{1}

\begin{widetext}

\section{Average Pauli weight}
We present the basic properties of the average Pauli weight $W(O)$ and provide reconstructed proofs based on Ref.~\cite{garcia2023resource}.

\begin{proof}[Proof: Faithfulness]
By definition of the average Pauli weight, $W(O)\geq0$ holds trivially.
Equality holds if and only if $O$ is the identity operator.
If $O$ is traceless, it contains no identity term in its Pauli expansion.
Thus, for any traceless operator, $W(O)\geq1$ holds, and $W(O)=1$ if and only if the Pauli expansion of $O$ consists solely of weight-1 Pauli operators.
\end{proof}

\begin{proof}[Proof: Invariance.]
    
We prove the invariance of the average Pauli weight: for nay operator $O$, if $U$ is a non-entangling unitary, then $W(U^\dagger OU) = W(O)$.
Here, a non-entangling unitary refers to any unitary that can be generated by single-qubit unitaries and swap operations.

In this proof, we expand the operator as $O = \sum_{\vec{a}} c_{\vec{a}} P_{\vec{a}}$, where the index $j$ denotes the position of the local Pauli operator.
The average Pauli weight can then be computed as follows:
\begin{align}
    W(O) &= \sum_j \sum_{\vec{a} (P_j\neq I)} |c_{\vec{a}}|^2 \notag\\
    &= \sum_j \frac{1}{d^{2n}} \sum_{\vec{a} (P_j\neq I)} \left| \mathrm{Tr}[OP_{\vec{a}}] \right|^2 \notag \\
    &= \sum_j \left[ \frac{1}{d^{2n}}\sum_{\vec{a}}\left| \mathrm{Tr}[OP_{\vec{a}}] \right|^2 - \frac{1}{d^{2n}}\sum_{\vec{a}(P_j=I)}\left| \mathrm{Tr}[OP_{\vec{a}}] \right|^2 \right] \notag \\
    &= \sum_j \left[ \frac{1}{d^n}\mathrm{Tr}[O^\dagger O] - \frac{1}{d^{n+1}}\mathrm{Tr}\left[ (\mathrm{Tr}_j[O])^\dagger (\mathrm{Tr}_j[O]) \right] \right].
\end{align}
Here, $d=2$ and $\mathrm{Tr}_j$ denotes the partial trace over the $j$th qubit.
We note that the partial trace operation removes all terms except those corresponding to Pauli strings with $P_j = I$.
When $U$ is generated as a product of single-qubit unitaries, the same calculation yields the following:
\begin{align}
    W(U^\dagger OU) &= \sum_j \left[ \frac{1}{d^n}\mathrm{Tr}[(U^\dagger OU)^\dagger (U^\dagger OU)] - \frac{1}{d^{n+1}}\mathrm{Tr}\left[ (\mathrm{Tr}_j[U^\dagger OU])^\dagger (\mathrm{Tr}_j[U^\dagger OU]) \right] \right] \notag \\
    &= \sum_j \left[ \frac{1}{d^n}\mathrm{Tr}[O^\dagger O] - \frac{1}{d^{n+1}}\mathrm{Tr}\left[ (\mathrm{Tr}_j[O])^\dagger (\mathrm{Tr}_j[O]) \right] \right].
\end{align}
The second equality follows from the fact that any local unitary operation on the $j$th qubit preserves terms with $P_j=I$ and cannot generate such terms from those with $P_j\neq I$.
Thus, if $U$ can be written as a product of local unitary operations as $U = \otimes_j U_j$, then $W(U^\dagger OU) = W(O)$ holds.
In addition, since the Pauli weight is clearly preserved under swap operations, $W(O)$ is invariant under swaps as well.
Therefore, the average Pauli weight is invariant under any non-entangling unitary, which can be generated as a product of single-qubit unitaries and swap operations.
\end{proof}

\begin{proof}[Proof: Additivity.]
Additivity can be proved directly.
By definition, the weight of Pauli strings is additive: $W(P_1\otimes P_2) = W(P_1) + W(P_2)$.
Therefore, for arbitrary normalized operators $O_1 = \sum_i c_iP_i$ and $O_2 = \sum_j c_j'P_j'$, additivity follows from the following calculation:
\begin{align}
    W(O_1\otimes O_2) &= \sum_{i,j}|c_ic_j'|^2 W(P_i\otimes P_j') \notag \\
    &= \sum_i |c_i|^2W(P_i) + \sum_j |c_j'|^2W(P_j') \notag \\
    &= W(O_1) + W(O_2).
\end{align}
\end{proof}

\section{Quantum Fourier entropy}
We provide proofs of the basic properties of the quantum Fourier entropy $S_a(O)$.

\begin{proof}[Proof: Faithfulness.]
Faithfulness follows immediately from the definition of $S_a(O)$, namely, its range is from 0 to $\log_a(4^n)$.
In particular, when the parameter $a$ is set to 4, the maximum value of $S(O)$ is $n$.
The equality $S_a(O)=0$ holds if and only if $O$ is a Pauli operator, since this corresponds to the case where the probability distribution is concentrated on a single element.
\end{proof}

\begin{proof}[Proof: Invariance.]    
Invariance follows directly from the definition of Clifford unitaries.
If $U$ is a Clifford unitary and $P$ is a Pauli operator, then $U^\dagger PU$ is also a Pauli operator, and the mapping is injective.
Therefore, when $O=\sum_i c_iP_i$ is conjugated by a Clifford unitary, it transforms into $U^\dagger OU = \sum_i c_i'P_i$, where $\{c_i'\}_i$ is a permutation of the original coefficients $\{c_i\}_i$.
Since the value of the quantum Fourier entropy is invariant under permutations of the coefficients, it remains unchanged under any Clifford unitary transformation.
\end{proof}

\begin{proof}[Proof: Additivity.]
Additivity can be derived through a straightforward argument.
For normalized operators $O_1 = \sum_i c_iP_i$ and $O_2 = \sum_j c'_j P'_j$, their tensor product is expressed as $O_1\otimes O_2 = \sum_{i,j}c_ic'_jP_i\otimes P'_j$.
Here, $\{P_i\otimes P'_j\}_{ij}$ forms the Pauli group on the total Hilbert space, and the coefficients $\{c_ic'_j\}_{ij}$ satisfy the normalization condition $\sum_{ij} |c_ic'_j|^2=1$.
Therefore, we can obtain the following equality:
\begin{align}
    S_a(O_1\otimes O_2) &= -\sum_{i,j} |c_ic'_j|^2\log|c_ic'_j|^2 \notag \\
    &= -\sum_i |c_i|^2\log|c_i|^2 -\sum_j |c'_j|^2\log|c'_j|^2 \notag \\
    &= S_a(O_1) + S_a(O_2).
\end{align}
\end{proof}

\section{Operator complexity}
In this section, we derive the operator $O_\mathrm{Max}$ that maximizes the operator complexity $R_a(O)$. 
We also compute the specific value of $W(O_\mathrm{Max})$, $S_a(O_\mathrm{Max})$, and $R_a(O_\mathrm{Max})$.

First, note that the values of these measures are independent of the complex phases of the coefficients of $O$.
Thus, we define $p_i \coloneqq |c_i|^2$ and assume that the operator can be written as $O = \sum_i \sqrt{p_i}P_i$.
To derive $O_\mathrm{Max}$, we solve the following optimization problem:
\begin{equation}
    \mathrm{maximize} \;\; R_a(O) = \sum_i p_i \left\{ W(P_i) -\log_a p_i \right\}.  \;\;\; \left( \sum_ip_i = 1,\;\; p_i\geq 0 \right)
\end{equation}
We define the Lagrangian function $\mathcal{L} \coloneqq \sum_i p_i \{ W(P_i) -\log_a p_i \} - \lambda(\sum_ip_i -1)$.
Then, the optimal coefficients $\{p_i\}$ satisfy the following conditions:
\begin{equation}
\label{Lag_cond}
    \frac{\partial\mathcal{L}}{\partial p_i} = w -\log_a p_i - \log_ae -\lambda = 0 \;\;\;(w = W(P_i)).
\end{equation}
From Eq.~\eqref{Lag_cond}, we obtain $p_i = a^{w-\lambda}/e =: p_w$, where $e$ denotes the base of the natural logarithm.
We note that the number of Pauli operators satisfying $W(P_i) = w$ is given by $\tbinom{n}{w}\cdot 3^w$.
From the normalization condition, we have:
\begin{equation}
    \sum_i p_i = \sum_{w=0}^n \binom{n}{w} 3^w \frac{a^{w-\lambda}}{e} = 1.
\end{equation}
Therefore, we obtain
\begin{equation}
    a^{-\lambda} = \frac{e}{(1+3a)^n}, \;\;\;\;\; p_w = \frac{a^w}{(1+3a)^n}.
\end{equation}
Taking into account the complex phases, the operator $O_{\mathrm{Max}}$ can be written as
\begin{equation}
    O_{\mathrm{Max}} = \sum_{w=0}^n \sum_{W(P_i)=w} e^{i\theta_i} \sqrt{ 
\frac{a^w}{(1+3a)^n} }P_i.
\end{equation}

Next, we provide a proof of Eq.~\eqref{WOmax} presented in the main text.
The value of $W(O_{\mathrm{Max}})$ can be expressed as
\begin{equation}
\label{WOmax_expand}
    W(O_{\mathrm{Max}}) = \sum_{w=0}^n \binom{n}{w}3^wp_ww = \frac{1}{(1+3a)^n}\sum_{w=0}^n \binom{n}{w}(3a)^w w =: \frac{1}{(1+3a)^n}S_n.
\end{equation}
To derive Eq.~\eqref{WOmax}, we prove the following equality by mathematical induction:
\begin{equation}
\label{sn}
    S_n = \sum_{w=0}^n \binom{n}{w}(3a)^w w = 3an(1+3a)^{n-1}.
\end{equation}
First, when $n=1$, Eq.~\eqref{sn} clearly holds.
Next, assume that Eq.~\eqref{sn} holds for some natural number $n$.
Under this assummption, we have
\begin{align}
    (1+3a)S_n &= \sum_{w=0}^n \binom{n}{w}(3a)^ww + \sum_{w=0}^n \binom{n}{w}(3a)^{w+1}w \notag \\
    &= \sum_{w=1}^n \binom{n}{w}(3a)^ww + \sum_{w=1}^{n+1} \binom{n}{w-1} (3a)^w(w-1) \notag \\
    &= \sum_{w=1}^n \left\{\binom{n}{w} + \binom{n}{w-1}\right\}(3a)^ww + (3a)^{n+1}(n+1) - \sum_{w=1}^{n+1} \binom{n}{w-1}(3a)^w \notag \\
    &= \sum_{w=1}^n \binom{n+1}{w}(3a)^ww + (3a)^{n+1}(n+1) - \frac{1}{n+1} \sum_{w=1}^{n+1} \frac{(n+1)!}{w!(n+1-w)!}(3a)^ww \notag \\
    &= \sum_{w=1}^{n+1} \binom{n+1}{w}(3a)^ww - \frac{1}{n+1}\sum_{w=1}^{n+1} \binom{n+1}{w} (3a)^ww \notag \\
    &= \frac{n}{n+1}S_{n+1},
\end{align}
from which we obtain
\begin{equation}
\label{sn+1}
    S_{n+1} = \frac{n+1}{n}(1+3a)S_n = 3a(n+1)(1+3a)^n.
\end{equation}
Eq.~\eqref{sn+1} shows that if Eq.~\eqref{sn} holds for some natural number $n$, it also holds for the next integer $n+1$.
Therefore, by mathematical induction, Eq.~\eqref{sn} holds for all positive integers $n$.
Using Eq.~\eqref{WOmax_expand}, we obtain
\begin{equation}
\label{WOmax_app}
    W(O_\mathrm{Max}) = \frac{3a}{1+3a}n.
\end{equation}

The value of $S(O_\mathrm{Max})$ can be obtained through a similar calculation:
\begin{align}
\label{SOmax_app}
    S(O_\mathrm{Max}) &= -\sum_{w=0}^n \binom{n}{w} 3^w \frac{a^w}{(1+3a)^n} \log_a \frac{a^w}{(1+3a)^n} \notag \\
    &= -\frac{1}{(1+3a)^n} \sum_{w=0}^n \binom{n}{w} (3a)^w \left\{w - n\log_a(1+3a) \right\} \notag \\
    &= -\frac{1}{(1+3a)^n} \left\{ S_n - (1+3a)^nn\log_a(1+3a) \right\} \notag \\
    &= n\log_a(1+3a) - \frac{3a}{1+3a}n.
\end{align}

From Eqs.~\eqref{WOmax_app} and \eqref{SOmax_app}, we obtain
\begin{equation}
    R(O_\mathrm{Max}) = W(O_\mathrm{Max}) + S(O_\mathrm{Max}) = n\log_a(1+3a).
\end{equation}

\section{Entanglement growth}
In this section, we provide proofs of the fundamental properties of the entanglement growth $G_E(U)$.
The overall structure of the proofs follows Ref.~\cite{bu2024complexity}.
However, since the invariance property stated in Ref.~\cite{bu2024complexity} contains an error, we correct it and provide a revised proof.

\begin{proof}[Proof: Faithfulness.]
The inequality $G_E(U) \geq 0$ follows from the faithfulness of the average Pauli weight.
For any unitary $U$ and a weight-1 Pauli operator $O$, we have
\begin{equation}
    G_E(U) \geq W(U^\dagger OU) - W(O) = W(U^\dagger OU)-1 \geq 0.
\end{equation}
Moreover, from the invariance of $W(O)$, it follows that when $U$ is a non-entangling unitary (i.e., generated from single-qubit unitaries and swap operations), we have equality $G_E(U)=0$.
Next, we show that any unitary satisfying $G_E(U)=0$ can be written as a product of single-qubit unitaries and swap operations.
By definition, such a unitary satisfies the relation $W(U^\dagger OU) = W(O)$ for any operator $O$.
To proceed, we consider $O = X_1$, where the subscript denotes the qubit on which the Pauli operator acts.
From the condition $W(U^\dagger X_1U) = W(X_1) = 1$, the following must hold:
\begin{equation}
\label{expand1}
    U^\dagger X_1U = \sum_i\alpha_iX_i + \sum_i\beta_iY_i + \sum_i\gamma_iZ_i,
\end{equation}
that is, $U^\dagger X_1U$ must be expressed solely in terms of weight-1 Pauli operators.
Suppose, for contradiction, that there exists a pair of indices $(i,j)$ such that $\alpha_i \neq 0, \alpha_j \neq 0,$ and $i\neq j$.
Then, squaring the right-hand side of Eq.~\eqref{expand1} yields a term of the form $2\alpha_i\alpha_j(X_i\otimes X_j)$.
However, this contradicts the identity $(U^\dagger X_1U)^2 = I$.
Therefore, Eq.~\eqref{expand1} must be rewritten as
\begin{equation}
\label{expand2}
    U^\dagger X_1U = \alpha_iX_i + \beta_jY_j + \gamma_k Z_k.
\end{equation}
Additionally, if $i\neq j$, squaring the right-hand side of Eq.~\eqref{expand2} yields a term of the form $2\alpha_i\beta_j(X_i\otimes Y_j)$.
Hence, Eq.~\eqref{expand2} must be further rewritten as
\begin{equation}
\label{expand3}
    U^\dagger X_1U = \alpha_iX_i + \beta_iY_i + \gamma_i Z_i.
\end{equation}
Similarly, when we take $O=Z_1$, we can write
\begin{equation}
\label{expand4}
    U^\dagger Z_1U = \alpha'_jX_j + \beta'_jY_j + \gamma'_j Z_j.
\end{equation}
If $i\neq j$, then $[U^\dagger X_1U, U^\dagger Z_1U] =0$ must hold.
However, this would imply $[X_1, Z_1]=0$, which is a contradiction.
Therefore, we must have $i=j$.
The same argument applies to the expansion of $U^\dagger Y_1U$.
Hence, for any local operator $A_1$, there exists a local unitary $V_1$ such that
\begin{align}
    U^\dagger(A_1\otimes I_{n-1})U &= (V_1^\dagger A_iV_1) \otimes I_{n-1} \notag \\
    (V_1\mathrm{SWAP}_{1,i}U)^\dagger(A_1\otimes& I_{n-1})(V_1\mathrm{SWAP}_{1,i}U) = A_1\otimes I_{n-1}.
\end{align}
From this relation, we can write $V_1\mathrm{SWAP}_{1,i}U = I_1\otimes V_{n-1}$, where $V_{n-1}$ is a unitary operator acting on the remaining $n-1$ qubits.
By iterating this process, any unitary $U$ satisfying $G_E(U)=0$ can be decomposed into a product of single-qubit unitaries and swap operations, as expressed in
\begin{equation}
    U = (\mathrm{SWAP}_{1,i}V_1)(\mathrm{SWAP}_{2,j}V_2)\cdots.
\end{equation}
Therefore, the necessary and sufficient condition for $G_E(U)=0$ is that $U$ is a non-entangling unitary.

\end{proof}

\begin{proof}[Proof: Invariance.]
For any non-entangling unitaries $V_1$ and $V_2$, the invariance of the entanglement growth follows from the following calculation:
\begin{align}
    G_E(V_1UV_2) &= \max_O \left\{ W(V_2^\dagger U^\dagger V_1^\dagger OV_1UV_2) - W(O) \right\} \notag \\
    &= \max_O \left\{ W(U^\dagger V_1^\dagger OV_1U) - W(O) \right\} \notag \\
    &= \max_O \left\{ W(U^\dagger OU) - W(V_1OV_1^\dagger) \right\} \notag \\
    &= \max_O \left\{ W(U^\dagger OU) - W(O) \right\} \notag \\
    &= G_E(U).
\end{align}
The second and fourth equalities hold due to the invariance of the average Pauli weight.
\end{proof}

\begin{proof}[Proof: Subadditivity]

We now prove the subadditivity $G_E(U_1\otimes U_2) \leq G_E(U_1) + G_E(U_2)$.
As a preliminary step, we first show that $G_E(UV) \leq G_E(U) + G_E(V)$, as follows:
\begin{align}
\label{ge1}
    G_E(UV) &= \max_O \left\{ W(V^\dagger U^\dagger OUV) - W(O) \right\} \notag \\
    &= \max_O \left\{ W(V^\dagger U^\dagger OUV) - W(U^\dagger OU) + W(U^\dagger OU) - W(O) \right\} \notag \\
    &\leq \max_O \left\{ W(V^\dagger U^\dagger OUV) - W(U^\dagger OU) \right\} + \max_O \left\{ W(U^\dagger OU) - W(O) \right\} \notag \\
    &= G_E(U) + G_E(V).
\end{align}
Next, we prove the inequality $G_E(U\otimes I_2) \leq G_E(U)$.
Let us consider the Hilbert space $\mathcal{H} = \mathcal{H}_1\otimes\mathcal{H}_2$, where $\mathcal{H}_1$ and $\mathcal{H}_2$ are $k$-qubit and $(n-k)$-qubit subsystems, respectively.
We expand an arbitrary operator $O$ as $O = \sum_{i,j}c_{ij}P_i\otimes P'_j$, where $\{P_i\}_i$ and $\{P_j'\}_j$ denote the Pauli bases on $\mathcal{H}_1$ and $\mathcal{H}_2$, respectively.
From the operator $O$, we define the following operators:
\begin{equation}
    O_j := \frac{1}{2^{n-k}}\mathrm{Tr}_2[O(I_1\otimes O_j)],\;\;\; B_j := \frac{O_j}{\|O_j\|},
\end{equation}
\begin{equation}
    O_i := \frac{1}{2^k}\mathrm{Tr}_1[O(P_i\otimes I_2)],\;\;\; A_i := \frac{O_i}{\|O_i\|}.
\end{equation}
Note that each $O_i$ acts on $\mathcal{H}_2$ and each $O_j$ acts on $\mathcal{H}_1$, and that $A_i$ and $B_j$ are the normalized versions of $O_i$ and $O_j$, respectively.
Since $O_j =\sum_ic_{ij}P_i$ and $O_i = \sum_jc_{ij}P'_j$, the average Pauli weight of $O$ can be expressed as
\begin{align}
\label{wexpand1}
    W(O) &= \sum_{i,j}|c_{ij}|^2 \left\{ W(P_i) + W(P'_j) \right\} \notag \\
    &= \sum_j\|O_j\|^2W(B_j) + \sum_i\|O_i\|^2W(A_i).
\end{align}
By applying the same argument to $(U^\dagger\otimes I_2)O(U\otimes I_2)$, we obtain
\begin{equation}
\label{wexpand2}
    W((U^\dagger\otimes I_2)O(U\otimes I_2)) = \sum_j\|O_j\|^2W(U^\dagger B_jU) + \sum_i\|O_i\|^2W(A_i).
\end{equation}
Combining Eqs.~\eqref{wexpand1} and \eqref{wexpand2}, we then arrive at
\begin{align}
    W((U^\dagger\otimes I_2)O(U\otimes I_2)) - W(O) &= \sum_j\|O_j\|^2 \left\{ W(U^\dagger B_jU) - W(O) \right\} \notag \\
    &\leq \max_{\|B_j\|=1} \left\{ W(U^\dagger B_jU) - W(O) \right\} \notag \\
    &= G_E(U).
\end{align}
Since this inequality holds for any operator $O$, we obtain
\begin{equation}
\label{ge2}
    G_E(U\otimes I_2) \leq G_E(U).
\end{equation}
Combining Eq.~\eqref{ge1}, Eq.~\eqref{ge2}, and similarly derived inequality $G_E(I_1\otimes U) \leq G_E(U)$, we conclude that for any unitary $U_1$ acting on $\mathcal{H_1}$ and $U_2$ acting on $\mathcal{H}_2$, the entanglement growth is subadditive:
\begin{equation}
    G_E(U_1) \leq G_E(U_1) + G_E(U_2).
\end{equation}

\end{proof}

\section{Nonstabilizerness growth}
In this section, we prove the basic properties of the nonstabilizerness growth $G_M(U)$.
The general structure of the proofs follows Ref.~\cite{bu2024complexity}, but we correct the condition for invariance stated therein, as it contains an inaccuracy.

\begin{proof}[Proof: Faithfulness.]
Faithfulness $G_M(U) \geq 0$ follows directly from the faithfulness of the quantum Fourier entropy.
For any unitary $U$, if $O$ is a Pauli operator, then
\begin{equation}
    G_M(U) \geq S_a(U^\dagger OU) -S(O) = S(U^\dagger OU) \geq 0.
\end{equation}
If $U$ is a Clifford unitary, then $G_M(U)=0$ due to the invariance of $S_a(O)$.
Conversely, if $U$ is a non-Clifford unitary, there exists a Pauli operator $O$ such that $U^\dagger OU$ is not a Pauli operator.
This implies $S_a(U^\dagger OU)-S(O)>0$, and hence $G_M(U)>0$.
Therefore, $G_M(U)=0$ if and only if $U$ is a Clifford unitary.
\end{proof}

\begin{proof}[Proof: Invariance.]
Invariance of the nonstabilizerness growth can be established as follows.
When both $V_1$ and $V_2$ are Clifford unitaries, we have
\begin{align}
    G_M(V_1UV_2) &= \max_O \left\{ S(V_2^\dagger U^\dagger V_1^\dagger OV_1UV_2) - S(O) \right\} \notag \\
    &= \max_O \left\{ S(U^\dagger V_1^\dagger OV_1U) - S(O) \right\} \notag \\
    &= \max_O \left\{ S(U^\dagger OU) - S(V_1OV_1^\dagger) \right\} \notag \\
    &= \max_O \left\{ S(U^\dagger OU) - S(O) \right\} \notag \\
    &= G_M(U).
\end{align}
The second and fourth equalities follow from the invariance of the quantum Fourier entropy.
\end{proof}

\begin{proof}[Proof: Subadditivity.]
We prove the subadditivity $G_M(U_1\otimes U_2) \leq G_M(U_1) + G_M(U_2)$.
The overall structure of the proof follows that of the entanglement growth case.
As a first step, we show that the inequality $G_M(UV) \leq G_M(U) + G_M(V)$ holds as follows:
\begin{align}
\label{gm1}
    G_M(UV) &= \max_O \left\{ S(V^\dagger U^\dagger OUV) - S(O) \right\} \notag \\
    &= \max_O \left\{ S(V^\dagger U^\dagger OUV) - S(U^\dagger OU) + S(U^\dagger OU) - S(O) \right\} \notag \\
    &\leq \max_O \left\{ S(V^\dagger U^\dagger OUV) - S(U^\dagger OU) \right\} + \max_O \left\{ S(U^\dagger OU) - S(O) \right\} \notag \\
    &= G_M(U) + G_M(V).
\end{align}
Next, we show the inequality $G_M(U\otimes I_2) \leq G_M(U)$.
Let us consider a subsystem consisting of $k$ qubits.
Given an operator $O = \sum_{i,j}c_{ij}P_i\otimes P'_j$ on the total system, we define the following operators on the subsystem:
\begin{equation}
    O_j := \frac{1}{2^{n-k}}\mathrm{Tr}_2[O(I_1\otimes O_j)],\;\;\; B_j := \frac{O_j}{\|O_j\|}.
\end{equation}
We note that $B_j = \sum_i c_{ij}/\|O_j\|P_i$.
Using this operator, we can compute $S(O)$ as
\begin{align}
\label{sexpand1}
    S(O) &= -\sum_{i,j}|c_{ij}|^2\log|c_{ij}|^2 \notag \\
    &= \sum_j \|O_j\|^2 S(B_j) - \sum_j\|O_j\|^2\log\|O_j\|^2.
\end{align}
By the same calculation, we obtain
\begin{equation}
\label{sexpand2}
    S((U^\dagger\otimes I_2)O(U\otimes I_2)) = \sum_j \|O_j\|^2 S(U^\dagger B_jU) - \sum_j\|O_j\|^2\log\|O_j\|^2.
\end{equation}
From Eqs.~\eqref{sexpand1} and \eqref{sexpand2}, the following inequality holds:
\begin{align}
    S((U^\dagger\otimes I_2)O(U\otimes I_2)) - S(O) &= \sum_j \|O_j\|^2 \left\{ S(U^\dagger B_j U) - S(B_j) \right\} \notag \\
    &\leq \max_{\|B_j\| = 1} \left\{ S(U^\dagger B_j U) - S(B_j) \right\} \notag \\
    &= G_M(U).
\end{align}
Since this inequality holds for any operator $O$, we obtain
\begin{equation}
\label{gm2}
    G_M(U\otimes I_2) \leq G_M(U).
\end{equation}
From Eqs.~\eqref{gm1} and \eqref{gm2}, together with the similarly derived inequality $G_M(I_1\otimes U)\leq G_M(U)$, we arrive at the subadditivity of the nonstabilizerness growth:
\begin{equation}
    G_M(U_1\otimes U_2) \leq G_M(U_1) + G_M(U_2).
\end{equation}

\end{proof}

\section{Complexity growth}
In this section, we prove the fundamental properties of the complexity growth.
The structure of the proofs is analogous to those of the entanglement growth and the nonstabilizerness growth.
In addition, we provide a proof of maxitivity, which is a distinctive property of $\tilde{G}(U)$.

\begin{proof}[Proof: Faithfulness.]
The inequality $G(U)\geq 0$ follows directly from the faithfulness of the operator complexity.
For any unitary $U$, if $O$ is a weight-1 Pauli operator, then
\begin{equation}
    G(U) \geq R(U^\dagger OU) - R(O) = R(U^\dagger OU) -1 \geq 0.
\end{equation}
If $U$ is a non-scrambling unitary, then $G(U)=0$ due to the invariance of $R(O)$.
Note that any non-scrambling unitary belongs to both the set of non-entangling unitaries and the set of Clifford unitaries.
Conversely, if $U$ is not a non-entangling unitary, there exists a weight-1 Pauli operator $O$ such that $W(U^\dagger OU) - 1 + S(U^\dagger OU) > 0$, which implies $G(U)>0$ (see the proof of the faithfulness of $W(O)$).
Similarly, if $U$ is a non-Clifford unitary, there also exists a weight-1 Pauli operator $O$ such that the same inequality holds.
Therefore, $G(U)=0$ if and only if $U$ is a non-scrambling unitary.
By the same reasoning, the faithfulness of $\tilde{G}(U)$ can also be established.

\end{proof}

\begin{proof}[Proof: Invariance.]
Invariance can be established in the same manner as for $G_E(U)$ and $G_M(U)$.
When both $V_1$ and $V_2$ are non-scrambling unitaries, we have
\begin{align}
    G(V_1UV_2) &= \max_O \left\{ R(V_2^\dagger U^\dagger V_1^\dagger OV_1UV_2) - R(O)  \right\} \notag \\
    &= \max_O \left\{ R(U^\dagger V_1^\dagger OV_1U) - R(O) \right\} \notag \\
    &= \max_O \left\{ R(U^\dagger OU) - R(V_1OV_1^\dagger) \right\} \notag \\
    &= \max_O \left\{ R(U^\dagger OU) - R(O) \right\} \notag \\
    &= G(U).
\end{align}
The second and fourth equalities follow from the invariance of $R(O)$.
By the same calculation, we obtain $\tilde{G}(V_1UV_2) = \tilde{G}(U)$.

\end{proof}

\begin{proof}[Proof: Subadditivity.]
We prove the subadditivity of $G(U)$ following the same structure as in the proofs for the entanglement growth and the nonstabilizerness growth.
We begin by showing that the inequality $G(UV)\leq G(U)+G(V)$ holds for any unitaries $U$ and $V$:
\begin{align}
\label{cg1}
    G(UV) &= \max_O \left\{ R(V^\dagger U^\dagger OUV) - R(O) \right\} \notag \\
    &= \max_O \left\{ R(V^\dagger U^\dagger OUV) - R(U^\dagger OU) + R(U^\dagger OU) - R(O) \right\} \notag \\
    &\leq \max_O \left\{ R(V^\dagger U^\dagger OUV) - R(U^\dagger OU) \right\} + \max_O \left\{ R(U^\dagger OU) - R(O) \right\} \notag \\
    &= G(U) + G(V).
\end{align}
Next, we prove the inequality $G(U\otimes I_2)\leq G(U)$.
We expand an arbitrary operator $O$ as $O=\sum_{i,j}c_{ij}P_i\otimes P'_j$.
Following the definitions used in the proofs of the entanglement growth and the nonstabilizerness growth, we introduce the following operators:
\begin{equation}
    O_j := \frac{1}{2^{n-k}}\mathrm{Tr}_2[O(I_1\otimes O_j)],\;\;\; B_j := \frac{O_j}{\|O_j\|},
\end{equation}
\begin{equation}
    O_i := \frac{1}{2^k}\mathrm{Tr}_1[O(P_i\otimes I_2)],\;\;\; A_i := \frac{O_i}{\|O_i\|}.
\end{equation}
By a similar calculation to those in the previous proofs, we obtain the following expression:
\begin{align}
\label{cg2}
    R(O) &= W(O) + S(O) \notag \\
    &= \sum_j \|O_j\|^2 W(B_j) + \sum_i\|O_i\|^2W(A_i) + \sum_j\|O_j\|^2 S(B_j) - \sum_j \|O_j\|^2\log \|O_j\|^2.
\end{align}
Applying this calculation to the operator $((U^\dagger\otimes I_2)O(U\otimes I_2))$, we obtain
\begin{align}
\label{cg3}
    & R((U^\dagger\otimes I_2)O(U\otimes I_2)) \notag \\
    &= \sum_j \|O_j\|^2 W(U^\dagger B_jU) + \sum_i\|O_i\|^2W(A_i) + \sum_j\|O_j\|^2 S(U^\dagger B_jU) - \sum_j \|O_j\|^2\log \|O_j\|^2.
\end{align}
From Eqs.~\eqref{cg2} and \eqref{cg3}, we obtain
\begin{align}
    R((U^\dagger\otimes I_2)O(U\otimes I_2)) - R(O) &= \sum_j \|O_j\|^2 \left\{ W(U^\dagger OU) - W(O) + S(U^\dagger B_jU) - S(B_j) \right\} \notag \\
    &\leq \max_{\|B_j\|=1} \left\{ W(U^\dagger OU) + S(U^\dagger OU) - W(O) - S(O) \right\} \notag \\
    &= G(U).
\end{align}
Since this inequality holds for any operator $O$, we obtain
\begin{equation}
\label{cg4}
    G(U\otimes I_2) \leq G(U).
\end{equation}
Combining Eqs.~\eqref{cg1} and \eqref{cg4} with the similarly derived inequality $G(I_1\otimes U)\leq G(U)$, we complete the proof of the subadditivity of $G(U)$:
\begin{equation}
    G(U_1\otimes U_2) \leq G(U_1) + G(U_2).
\end{equation}

\end{proof}

\begin{proof}[Proof: Maxitivity.]
Maxitivity of $\tilde{G}(U)$ can be established through the following calculation:
\begin{align}
    \tilde{G}(U_1\otimes U_2) &= \max_{W(O)=1, S(O)=0} \left\{ R((U_1^\dagger\otimes U_2^\dagger) O(U_1\otimes U_2)) - 1 \right\} \notag \\
    &= \max\left\{  \max_{W(O_1)=1, S(O_1)=0}\{ R(U_1^\dagger OU_1)-1 \}, \max_{W(O_2)=1, S(O_2)=0}\{ R(U_2^\dagger OU_2)-1 \}    \right\} \notag \\
    &= \max\left\{ \tilde{G}(U_1),  \tilde{G}(U_2) \right\}.
\end{align}
Here, we note that an operator $O$ satisfies $W(O)=1$ and $S(O)=0$ if and only if $O$ is a weight-1 Pauli operator.
The second equality follows from the faithfulness and additivity of the operator complexity.

\end{proof}

\end{widetext}


\begin{thebibliography}{57}%
\makeatletter
\providecommand \@ifxundefined [1]{%
 \@ifx{#1\undefined}
}%
\providecommand \@ifnum [1]{%
 \ifnum #1\expandafter \@firstoftwo
 \else \expandafter \@secondoftwo
 \fi
}%
\providecommand \@ifx [1]{%
 \ifx #1\expandafter \@firstoftwo
 \else \expandafter \@secondoftwo
 \fi
}%
\providecommand \natexlab [1]{#1}%
\providecommand \enquote  [1]{``#1''}%
\providecommand \bibnamefont  [1]{#1}%
\providecommand \bibfnamefont [1]{#1}%
\providecommand \citenamefont [1]{#1}%
\providecommand \href@noop [0]{\@secondoftwo}%
\providecommand \href [0]{\begingroup \@sanitize@url \@href}%
\providecommand \@href[1]{\@@startlink{#1}\@@href}%
\providecommand \@@href[1]{\endgroup#1\@@endlink}%
\providecommand \@sanitize@url [0]{\catcode `\\12\catcode `\$12\catcode `\&12\catcode `\#12\catcode `\^12\catcode `\_12\catcode `\%12\relax}%
\providecommand \@@startlink[1]{}%
\providecommand \@@endlink[0]{}%
\providecommand \url  [0]{\begingroup\@sanitize@url \@url }%
\providecommand \@url [1]{\endgroup\@href {#1}{\urlprefix }}%
\providecommand \urlprefix  [0]{URL }%
\providecommand \Eprint [0]{\href }%
\providecommand \doibase [0]{https://doi.org/}%
\providecommand \selectlanguage [0]{\@gobble}%
\providecommand \bibinfo  [0]{\@secondoftwo}%
\providecommand \bibfield  [0]{\@secondoftwo}%
\providecommand \translation [1]{[#1]}%
\providecommand \BibitemOpen [0]{}%
\providecommand \bibitemStop [0]{}%
\providecommand \bibitemNoStop [0]{.\EOS\space}%
\providecommand \EOS [0]{\spacefactor3000\relax}%
\providecommand \BibitemShut  [1]{\csname bibitem#1\endcsname}%
\let\auto@bib@innerbib\@empty
\bibitem [{\citenamefont {Horodecki}\ \emph {et~al.}(2009)\citenamefont {Horodecki}, \citenamefont {Horodecki}, \citenamefont {Horodecki},\ and\ \citenamefont {Horodecki}}]{horodecki2009quantum}%
  \BibitemOpen
  \bibfield  {author} {\bibinfo {author} {\bibfnamefont {R.}~\bibnamefont {Horodecki}}, \bibinfo {author} {\bibfnamefont {P.}~\bibnamefont {Horodecki}}, \bibinfo {author} {\bibfnamefont {M.}~\bibnamefont {Horodecki}},\ and\ \bibinfo {author} {\bibfnamefont {K.}~\bibnamefont {Horodecki}},\ }\bibfield  {title} {\bibinfo {title} {Quantum entanglement},\ }\href {https://doi.org/10.1103/RevModPhys.81.865} {\bibfield  {journal} {\bibinfo  {journal} {Rev. Mod. Phys.}\ }\textbf {\bibinfo {volume} {81}},\ \bibinfo {pages} {865} (\bibinfo {year} {2009})}\BibitemShut {NoStop}%
\bibitem [{\citenamefont {Bennett}\ \emph {et~al.}(1996)\citenamefont {Bennett}, \citenamefont {Bernstein}, \citenamefont {Popescu},\ and\ \citenamefont {Schumacher}}]{bennett1996concentrating}%
  \BibitemOpen
  \bibfield  {author} {\bibinfo {author} {\bibfnamefont {C.~H.}\ \bibnamefont {Bennett}}, \bibinfo {author} {\bibfnamefont {H.~J.}\ \bibnamefont {Bernstein}}, \bibinfo {author} {\bibfnamefont {S.}~\bibnamefont {Popescu}},\ and\ \bibinfo {author} {\bibfnamefont {B.}~\bibnamefont {Schumacher}},\ }\bibfield  {title} {\bibinfo {title} {Concentrating partial entanglement by local operations},\ }\href {https://doi.org/10.1103/PhysRevA.53.2046} {\bibfield  {journal} {\bibinfo  {journal} {Phys. Rev. A}\ }\textbf {\bibinfo {volume} {53}},\ \bibinfo {pages} {2046} (\bibinfo {year} {1996})}\BibitemShut {NoStop}%
\bibitem [{\citenamefont {Vedral}\ and\ \citenamefont {Plenio}(1998)}]{vedral1998entanglement}%
  \BibitemOpen
  \bibfield  {author} {\bibinfo {author} {\bibfnamefont {V.}~\bibnamefont {Vedral}}\ and\ \bibinfo {author} {\bibfnamefont {M.~B.}\ \bibnamefont {Plenio}},\ }\bibfield  {title} {\bibinfo {title} {Entanglement measures and purification procedures},\ }\href {https://doi.org/10.1103/PhysRevA.57.1619} {\bibfield  {journal} {\bibinfo  {journal} {Phys. Rev. A}\ }\textbf {\bibinfo {volume} {57}},\ \bibinfo {pages} {1619} (\bibinfo {year} {1998})}\BibitemShut {NoStop}%
\bibitem [{\citenamefont {Lami}\ and\ \citenamefont {Regula}(2023)}]{lami2023no}%
  \BibitemOpen
  \bibfield  {author} {\bibinfo {author} {\bibfnamefont {L.}~\bibnamefont {Lami}}\ and\ \bibinfo {author} {\bibfnamefont {B.}~\bibnamefont {Regula}},\ }\bibfield  {title} {\bibinfo {title} {No second law of entanglement manipulation after all},\ }\href {https://doi.org/10.1038/s41567-022-01873-9} {\bibfield  {journal} {\bibinfo  {journal} {Nature Physics}\ }\textbf {\bibinfo {volume} {19}},\ \bibinfo {pages} {184} (\bibinfo {year} {2023})}\BibitemShut {NoStop}%
\bibitem [{\citenamefont {Baumgratz}\ \emph {et~al.}(2014)\citenamefont {Baumgratz}, \citenamefont {Cramer},\ and\ \citenamefont {Plenio}}]{baumgratz2014quantifying}%
  \BibitemOpen
  \bibfield  {author} {\bibinfo {author} {\bibfnamefont {T.}~\bibnamefont {Baumgratz}}, \bibinfo {author} {\bibfnamefont {M.}~\bibnamefont {Cramer}},\ and\ \bibinfo {author} {\bibfnamefont {M.~B.}\ \bibnamefont {Plenio}},\ }\bibfield  {title} {\bibinfo {title} {Quantifying coherence},\ }\href {https://doi.org/10.1103/PhysRevLett.113.140401} {\bibfield  {journal} {\bibinfo  {journal} {Phys. Rev. Lett.}\ }\textbf {\bibinfo {volume} {113}},\ \bibinfo {pages} {140401} (\bibinfo {year} {2014})}\BibitemShut {NoStop}%
\bibitem [{\citenamefont {Winter}\ and\ \citenamefont {Yang}(2016)}]{winter2016operational}%
  \BibitemOpen
  \bibfield  {author} {\bibinfo {author} {\bibfnamefont {A.}~\bibnamefont {Winter}}\ and\ \bibinfo {author} {\bibfnamefont {D.}~\bibnamefont {Yang}},\ }\bibfield  {title} {\bibinfo {title} {Operational resource theory of coherence},\ }\href {https://doi.org/10.1103/PhysRevLett.116.120404} {\bibfield  {journal} {\bibinfo  {journal} {Phys. Rev. Lett.}\ }\textbf {\bibinfo {volume} {116}},\ \bibinfo {pages} {120404} (\bibinfo {year} {2016})}\BibitemShut {NoStop}%
\bibitem [{\citenamefont {Marvian}\ \emph {et~al.}(2016)\citenamefont {Marvian}, \citenamefont {Spekkens},\ and\ \citenamefont {Zanardi}}]{marvian2016quantum}%
  \BibitemOpen
  \bibfield  {author} {\bibinfo {author} {\bibfnamefont {I.}~\bibnamefont {Marvian}}, \bibinfo {author} {\bibfnamefont {R.~W.}\ \bibnamefont {Spekkens}},\ and\ \bibinfo {author} {\bibfnamefont {P.}~\bibnamefont {Zanardi}},\ }\bibfield  {title} {\bibinfo {title} {Quantum speed limits, coherence, and asymmetry},\ }\href {https://doi.org/10.1103/PhysRevA.93.052331} {\bibfield  {journal} {\bibinfo  {journal} {Phys. Rev. A}\ }\textbf {\bibinfo {volume} {93}},\ \bibinfo {pages} {052331} (\bibinfo {year} {2016})}\BibitemShut {NoStop}%
\bibitem [{\citenamefont {Napoli}\ \emph {et~al.}(2016)\citenamefont {Napoli}, \citenamefont {Bromley}, \citenamefont {Cianciaruso}, \citenamefont {Piani}, \citenamefont {Johnston},\ and\ \citenamefont {Adesso}}]{napoli2016robustness}%
  \BibitemOpen
  \bibfield  {author} {\bibinfo {author} {\bibfnamefont {C.}~\bibnamefont {Napoli}}, \bibinfo {author} {\bibfnamefont {T.~R.}\ \bibnamefont {Bromley}}, \bibinfo {author} {\bibfnamefont {M.}~\bibnamefont {Cianciaruso}}, \bibinfo {author} {\bibfnamefont {M.}~\bibnamefont {Piani}}, \bibinfo {author} {\bibfnamefont {N.}~\bibnamefont {Johnston}},\ and\ \bibinfo {author} {\bibfnamefont {G.}~\bibnamefont {Adesso}},\ }\bibfield  {title} {\bibinfo {title} {Robustness of coherence: an operational and observable measure of quantum coherence},\ }\href {https://doi.org/10.1103/PhysRevLett.116.150502} {\bibfield  {journal} {\bibinfo  {journal} {Phys. Rev. Lett.}\ }\textbf {\bibinfo {volume} {116}},\ \bibinfo {pages} {150502} (\bibinfo {year} {2016})}\BibitemShut {NoStop}%
\bibitem [{\citenamefont {Saxena}\ \emph {et~al.}(2020)\citenamefont {Saxena}, \citenamefont {Chitambar},\ and\ \citenamefont {Gour}}]{saxena2020dynamical}%
  \BibitemOpen
  \bibfield  {author} {\bibinfo {author} {\bibfnamefont {G.}~\bibnamefont {Saxena}}, \bibinfo {author} {\bibfnamefont {E.}~\bibnamefont {Chitambar}},\ and\ \bibinfo {author} {\bibfnamefont {G.}~\bibnamefont {Gour}},\ }\bibfield  {title} {\bibinfo {title} {Dynamical resource theory of quantum coherence},\ }\href {https://doi.org/10.1103/PhysRevResearch.2.023298} {\bibfield  {journal} {\bibinfo  {journal} {Phys. Rev. Res.}\ }\textbf {\bibinfo {volume} {2}},\ \bibinfo {pages} {023298} (\bibinfo {year} {2020})}\BibitemShut {NoStop}%
\bibitem [{\citenamefont {Shiraishi}\ and\ \citenamefont {Takagi}(2024)}]{shiraishi2024arbitrary}%
  \BibitemOpen
  \bibfield  {author} {\bibinfo {author} {\bibfnamefont {N.}~\bibnamefont {Shiraishi}}\ and\ \bibinfo {author} {\bibfnamefont {R.}~\bibnamefont {Takagi}},\ }\bibfield  {title} {\bibinfo {title} {Arbitrary amplification of quantum coherence in asymptotic and catalytic transformation},\ }\href {https://doi.org/10.1103/PhysRevLett.132.180202} {\bibfield  {journal} {\bibinfo  {journal} {Phys. Rev. Lett.}\ }\textbf {\bibinfo {volume} {132}},\ \bibinfo {pages} {180202} (\bibinfo {year} {2024})}\BibitemShut {NoStop}%
\bibitem [{\citenamefont {Theurer}\ \emph {et~al.}(2017)\citenamefont {Theurer}, \citenamefont {Killoran}, \citenamefont {Egloff},\ and\ \citenamefont {Plenio}}]{theurer2017resource}%
  \BibitemOpen
  \bibfield  {author} {\bibinfo {author} {\bibfnamefont {T.}~\bibnamefont {Theurer}}, \bibinfo {author} {\bibfnamefont {N.}~\bibnamefont {Killoran}}, \bibinfo {author} {\bibfnamefont {D.}~\bibnamefont {Egloff}},\ and\ \bibinfo {author} {\bibfnamefont {M.~B.}\ \bibnamefont {Plenio}},\ }\bibfield  {title} {\bibinfo {title} {Resource theory of superposition},\ }\href {https://doi.org/10.1103/PhysRevLett.119.230401} {\bibfield  {journal} {\bibinfo  {journal} {Phys. Rev. Lett.}\ }\textbf {\bibinfo {volume} {119}},\ \bibinfo {pages} {230401} (\bibinfo {year} {2017})}\BibitemShut {NoStop}%
\bibitem [{\citenamefont {Guff}\ \emph {et~al.}(2021)\citenamefont {Guff}, \citenamefont {McMahon}, \citenamefont {Sanders},\ and\ \citenamefont {Gilchrist}}]{guff2021resource}%
  \BibitemOpen
  \bibfield  {author} {\bibinfo {author} {\bibfnamefont {T.}~\bibnamefont {Guff}}, \bibinfo {author} {\bibfnamefont {N.~A.}\ \bibnamefont {McMahon}}, \bibinfo {author} {\bibfnamefont {Y.~R.}\ \bibnamefont {Sanders}},\ and\ \bibinfo {author} {\bibfnamefont {A.}~\bibnamefont {Gilchrist}},\ }\bibfield  {title} {\bibinfo {title} {A resource theory of quantum measurements},\ }\href {https://doi.org/10.1088/1751-8121/abed67} {\bibfield  {journal} {\bibinfo  {journal} {J. Phys. A: Math. Theor.}\ }\textbf {\bibinfo {volume} {54}},\ \bibinfo {pages} {225301} (\bibinfo {year} {2021})}\BibitemShut {NoStop}%
\bibitem [{\citenamefont {Veitch}\ \emph {et~al.}(2014)\citenamefont {Veitch}, \citenamefont {Mousavian}, \citenamefont {Gottesman},\ and\ \citenamefont {Emerson}}]{veitch2014resource}%
  \BibitemOpen
  \bibfield  {author} {\bibinfo {author} {\bibfnamefont {V.}~\bibnamefont {Veitch}}, \bibinfo {author} {\bibfnamefont {S.~H.}\ \bibnamefont {Mousavian}}, \bibinfo {author} {\bibfnamefont {D.}~\bibnamefont {Gottesman}},\ and\ \bibinfo {author} {\bibfnamefont {J.}~\bibnamefont {Emerson}},\ }\bibfield  {title} {\bibinfo {title} {The resource theory of stabilizer quantum computation},\ }\href {https://doi.org/10.1088/1367-2630/16/1/013009} {\bibfield  {journal} {\bibinfo  {journal} {New J. Phys.}\ }\textbf {\bibinfo {volume} {16}},\ \bibinfo {pages} {013009} (\bibinfo {year} {2014})}\BibitemShut {NoStop}%
\bibitem [{\citenamefont {Howard}\ and\ \citenamefont {Campbell}(2017)}]{howard2017application}%
  \BibitemOpen
  \bibfield  {author} {\bibinfo {author} {\bibfnamefont {M.}~\bibnamefont {Howard}}\ and\ \bibinfo {author} {\bibfnamefont {E.}~\bibnamefont {Campbell}},\ }\bibfield  {title} {\bibinfo {title} {Application of a resource theory for magic states to fault-tolerant quantum computing},\ }\href {https://doi.org/10.1103/PhysRevLett.118.090501} {\bibfield  {journal} {\bibinfo  {journal} {Phys. Rev. Lett.}\ }\textbf {\bibinfo {volume} {118}},\ \bibinfo {pages} {090501} (\bibinfo {year} {2017})}\BibitemShut {NoStop}%
\bibitem [{\citenamefont {Leone}\ \emph {et~al.}(2022)\citenamefont {Leone}, \citenamefont {Oliviero},\ and\ \citenamefont {Hamma}}]{leone2022stabilizer}%
  \BibitemOpen
  \bibfield  {author} {\bibinfo {author} {\bibfnamefont {L.}~\bibnamefont {Leone}}, \bibinfo {author} {\bibfnamefont {S.~F.}\ \bibnamefont {Oliviero}},\ and\ \bibinfo {author} {\bibfnamefont {A.}~\bibnamefont {Hamma}},\ }\bibfield  {title} {\bibinfo {title} {Stabilizer {R}{\'e}nyi entropy},\ }\href {https://doi.org/10.1103/PhysRevLett.128.050402} {\bibfield  {journal} {\bibinfo  {journal} {Phys. Rev. Lett.}\ }\textbf {\bibinfo {volume} {128}},\ \bibinfo {pages} {050402} (\bibinfo {year} {2022})}\BibitemShut {NoStop}%
\bibitem [{\citenamefont {Haug}\ and\ \citenamefont {Piroli}(2023{\natexlab{a}})}]{haug2023stabilizer}%
  \BibitemOpen
  \bibfield  {author} {\bibinfo {author} {\bibfnamefont {T.}~\bibnamefont {Haug}}\ and\ \bibinfo {author} {\bibfnamefont {L.}~\bibnamefont {Piroli}},\ }\bibfield  {title} {\bibinfo {title} {Stabilizer entropies and nonstabilizerness monotones},\ }\href {https://doi.org/10.22331/q-2023-08-28-1092} {\bibfield  {journal} {\bibinfo  {journal} {Quantum}\ }\textbf {\bibinfo {volume} {7}},\ \bibinfo {pages} {1092} (\bibinfo {year} {2023}{\natexlab{a}})}\BibitemShut {NoStop}%
\bibitem [{\citenamefont {Haug}\ and\ \citenamefont {Piroli}(2023{\natexlab{b}})}]{haug2023quantifying}%
  \BibitemOpen
  \bibfield  {author} {\bibinfo {author} {\bibfnamefont {T.}~\bibnamefont {Haug}}\ and\ \bibinfo {author} {\bibfnamefont {L.}~\bibnamefont {Piroli}},\ }\bibfield  {title} {\bibinfo {title} {Quantifying nonstabilizerness of matrix product states},\ }\href {https://doi.org/10.1103/PhysRevB.107.035148} {\bibfield  {journal} {\bibinfo  {journal} {Phys. Rev. B}\ }\textbf {\bibinfo {volume} {107}},\ \bibinfo {pages} {035148} (\bibinfo {year} {2023}{\natexlab{b}})}\BibitemShut {NoStop}%
\bibitem [{\citenamefont {Leone}\ and\ \citenamefont {Bittel}(2024)}]{leone2024stabilizer}%
  \BibitemOpen
  \bibfield  {author} {\bibinfo {author} {\bibfnamefont {L.}~\bibnamefont {Leone}}\ and\ \bibinfo {author} {\bibfnamefont {L.}~\bibnamefont {Bittel}},\ }\bibfield  {title} {\bibinfo {title} {Stabilizer entropies are monotones for magic-state resource theory},\ }\href {https://doi.org/10.1103/PhysRevA.110.L040403} {\bibfield  {journal} {\bibinfo  {journal} {Phys. Rev. A}\ }\textbf {\bibinfo {volume} {110}},\ \bibinfo {pages} {L040403} (\bibinfo {year} {2024})}\BibitemShut {NoStop}%
\bibitem [{\citenamefont {Zhang}\ \emph {et~al.}(2024)\citenamefont {Zhang}, \citenamefont {Feng},\ and\ \citenamefont {Luo}}]{zhang2024no}%
  \BibitemOpen
  \bibfield  {author} {\bibinfo {author} {\bibfnamefont {Z.}~\bibnamefont {Zhang}}, \bibinfo {author} {\bibfnamefont {L.}~\bibnamefont {Feng}},\ and\ \bibinfo {author} {\bibfnamefont {S.}~\bibnamefont {Luo}},\ }\bibfield  {title} {\bibinfo {title} {No-broadcasting of magic states},\ }\href {https://doi.org/10.1103/PhysRevA.110.012462} {\bibfield  {journal} {\bibinfo  {journal} {Phys. Rev. A}\ }\textbf {\bibinfo {volume} {110}},\ \bibinfo {pages} {012462} (\bibinfo {year} {2024})}\BibitemShut {NoStop}%
\bibitem [{\citenamefont {Yunger~Halpern}\ \emph {et~al.}(2022)\citenamefont {Yunger~Halpern}, \citenamefont {Kothakonda}, \citenamefont {Haferkamp}, \citenamefont {Munson}, \citenamefont {Eisert},\ and\ \citenamefont {Faist}}]{yunger2022resource}%
  \BibitemOpen
  \bibfield  {author} {\bibinfo {author} {\bibfnamefont {N.}~\bibnamefont {Yunger~Halpern}}, \bibinfo {author} {\bibfnamefont {N.~B.}\ \bibnamefont {Kothakonda}}, \bibinfo {author} {\bibfnamefont {J.}~\bibnamefont {Haferkamp}}, \bibinfo {author} {\bibfnamefont {A.}~\bibnamefont {Munson}}, \bibinfo {author} {\bibfnamefont {J.}~\bibnamefont {Eisert}},\ and\ \bibinfo {author} {\bibfnamefont {P.}~\bibnamefont {Faist}},\ }\bibfield  {title} {\bibinfo {title} {Resource theory of quantum uncomplexity},\ }\href {https://doi.org/10.1103/PhysRevA.106.062417} {\bibfield  {journal} {\bibinfo  {journal} {Phys. Rev. A}\ }\textbf {\bibinfo {volume} {106}},\ \bibinfo {pages} {062417} (\bibinfo {year} {2022})}\BibitemShut {NoStop}%
\bibitem [{\citenamefont {Lostaglio}(2019)}]{lostaglio2019introductory}%
  \BibitemOpen
  \bibfield  {author} {\bibinfo {author} {\bibfnamefont {M.}~\bibnamefont {Lostaglio}},\ }\bibfield  {title} {\bibinfo {title} {An introductory review of the resource theory approach to thermodynamics},\ }\href {https://doi.org/10.1088/1361-6633/ab46e5} {\bibfield  {journal} {\bibinfo  {journal} {Rep. Prog. Phys.}\ }\textbf {\bibinfo {volume} {82}},\ \bibinfo {pages} {114001} (\bibinfo {year} {2019})}\BibitemShut {NoStop}%
\bibitem [{\citenamefont {Brandao}\ \emph {et~al.}(2013)\citenamefont {Brandao}, \citenamefont {Horodecki}, \citenamefont {Oppenheim}, \citenamefont {Renes},\ and\ \citenamefont {Spekkens}}]{brandao2013resource}%
  \BibitemOpen
  \bibfield  {author} {\bibinfo {author} {\bibfnamefont {F.~G.}\ \bibnamefont {Brandao}}, \bibinfo {author} {\bibfnamefont {M.}~\bibnamefont {Horodecki}}, \bibinfo {author} {\bibfnamefont {J.}~\bibnamefont {Oppenheim}}, \bibinfo {author} {\bibfnamefont {J.~M.}\ \bibnamefont {Renes}},\ and\ \bibinfo {author} {\bibfnamefont {R.~W.}\ \bibnamefont {Spekkens}},\ }\bibfield  {title} {\bibinfo {title} {Resource theory of quantum states out of thermal equilibrium},\ }\href {https://doi.org/10.1103/PhysRevLett.111.250404} {\bibfield  {journal} {\bibinfo  {journal} {Phys. Rev. Lett.}\ }\textbf {\bibinfo {volume} {111}},\ \bibinfo {pages} {250404} (\bibinfo {year} {2013})}\BibitemShut {NoStop}%
\bibitem [{\citenamefont {Liu}\ \emph {et~al.}(2019)\citenamefont {Liu}, \citenamefont {Bu},\ and\ \citenamefont {Takagi}}]{liu2019one}%
  \BibitemOpen
  \bibfield  {author} {\bibinfo {author} {\bibfnamefont {Z.-W.}\ \bibnamefont {Liu}}, \bibinfo {author} {\bibfnamefont {K.}~\bibnamefont {Bu}},\ and\ \bibinfo {author} {\bibfnamefont {R.}~\bibnamefont {Takagi}},\ }\bibfield  {title} {\bibinfo {title} {One-shot operational quantum resource theory},\ }\href {https://doi.org/10.1103/PhysRevLett.123.020401} {\bibfield  {journal} {\bibinfo  {journal} {Phys. Rev. Lett.}\ }\textbf {\bibinfo {volume} {123}},\ \bibinfo {pages} {020401} (\bibinfo {year} {2019})}\BibitemShut {NoStop}%
\bibitem [{\citenamefont {Regula}(2022)}]{regula2022probabilistic}%
  \BibitemOpen
  \bibfield  {author} {\bibinfo {author} {\bibfnamefont {B.}~\bibnamefont {Regula}},\ }\bibfield  {title} {\bibinfo {title} {Probabilistic transformations of quantum resources},\ }\href {https://doi.org/10.1103/PhysRevLett.128.110505} {\bibfield  {journal} {\bibinfo  {journal} {Phys. Rev. Lett.}\ }\textbf {\bibinfo {volume} {128}},\ \bibinfo {pages} {110505} (\bibinfo {year} {2022})}\BibitemShut {NoStop}%
\bibitem [{\citenamefont {Zanoni}\ and\ \citenamefont {Scandolo}(2025)}]{zanoni2025choi}%
  \BibitemOpen
  \bibfield  {author} {\bibinfo {author} {\bibfnamefont {E.}~\bibnamefont {Zanoni}}\ and\ \bibinfo {author} {\bibfnamefont {C.~M.}\ \bibnamefont {Scandolo}},\ }\bibfield  {title} {\bibinfo {title} {Choi-defined resource theories},\ }\href {https://doi.org/10.1103/PhysRevA.111.062407} {\bibfield  {journal} {\bibinfo  {journal} {Phys. Rev. A}\ }\textbf {\bibinfo {volume} {111}},\ \bibinfo {pages} {062407} (\bibinfo {year} {2025})}\BibitemShut {NoStop}%
\bibitem [{\citenamefont {Chitambar}\ and\ \citenamefont {Gour}(2019)}]{chitambar2019quantum}%
  \BibitemOpen
  \bibfield  {author} {\bibinfo {author} {\bibfnamefont {E.}~\bibnamefont {Chitambar}}\ and\ \bibinfo {author} {\bibfnamefont {G.}~\bibnamefont {Gour}},\ }\bibfield  {title} {\bibinfo {title} {Quantum resource theories},\ }\href {https://doi.org/10.1103/RevModPhys.91.025001} {\bibfield  {journal} {\bibinfo  {journal} {Rev. Mod. Phys.}\ }\textbf {\bibinfo {volume} {91}},\ \bibinfo {pages} {025001} (\bibinfo {year} {2019})}\BibitemShut {NoStop}%
\bibitem [{\citenamefont {Gu}\ \emph {et~al.}(2025)\citenamefont {Gu}, \citenamefont {Oliviero},\ and\ \citenamefont {Leone}}]{gu2025magic}%
  \BibitemOpen
  \bibfield  {author} {\bibinfo {author} {\bibfnamefont {A.}~\bibnamefont {Gu}}, \bibinfo {author} {\bibfnamefont {S.~F.}\ \bibnamefont {Oliviero}},\ and\ \bibinfo {author} {\bibfnamefont {L.}~\bibnamefont {Leone}},\ }\bibfield  {title} {\bibinfo {title} {Magic-induced computational separation in entanglement theory},\ }\href {https://doi.org/10.1103/PRXQuantum.6.020324} {\bibfield  {journal} {\bibinfo  {journal} {PRX Quantum}\ }\textbf {\bibinfo {volume} {6}},\ \bibinfo {pages} {020324} (\bibinfo {year} {2025})}\BibitemShut {NoStop}%
\bibitem [{\citenamefont {Dowling}\ \emph {et~al.}(2025{\natexlab{a}})\citenamefont {Dowling}, \citenamefont {Modi},\ and\ \citenamefont {White}}]{dowling2025bridging}%
  \BibitemOpen
  \bibfield  {author} {\bibinfo {author} {\bibfnamefont {N.}~\bibnamefont {Dowling}}, \bibinfo {author} {\bibfnamefont {K.}~\bibnamefont {Modi}},\ and\ \bibinfo {author} {\bibfnamefont {G.~A.}\ \bibnamefont {White}},\ }\bibfield  {title} {\bibinfo {title} {Bridging entanglement and magic resources within operator space},\ }\href {https://doi.org/10.1103/c7k1-xcwy} {\bibfield  {journal} {\bibinfo  {journal} {Phys. Rev. Lett.}\ }\textbf {\bibinfo {volume} {135}},\ \bibinfo {pages} {160201} (\bibinfo {year} {2025}{\natexlab{a}})}\BibitemShut {NoStop}%
\bibitem [{\citenamefont {Tirrito}\ \emph {et~al.}(2024)\citenamefont {Tirrito}, \citenamefont {Tarabunga}, \citenamefont {Lami}, \citenamefont {Chanda}, \citenamefont {Leone}, \citenamefont {Oliviero}, \citenamefont {Dalmonte}, \citenamefont {Collura},\ and\ \citenamefont {Hamma}}]{tirrito2024quantifying}%
  \BibitemOpen
  \bibfield  {author} {\bibinfo {author} {\bibfnamefont {E.}~\bibnamefont {Tirrito}}, \bibinfo {author} {\bibfnamefont {P.~S.}\ \bibnamefont {Tarabunga}}, \bibinfo {author} {\bibfnamefont {G.}~\bibnamefont {Lami}}, \bibinfo {author} {\bibfnamefont {T.}~\bibnamefont {Chanda}}, \bibinfo {author} {\bibfnamefont {L.}~\bibnamefont {Leone}}, \bibinfo {author} {\bibfnamefont {S.~F.}\ \bibnamefont {Oliviero}}, \bibinfo {author} {\bibfnamefont {M.}~\bibnamefont {Dalmonte}}, \bibinfo {author} {\bibfnamefont {M.}~\bibnamefont {Collura}},\ and\ \bibinfo {author} {\bibfnamefont {A.}~\bibnamefont {Hamma}},\ }\bibfield  {title} {\bibinfo {title} {Quantifying nonstabilizerness through entanglement spectrum flatness},\ }\href {https://doi.org/10.1103/PhysRevA.109.L040401} {\bibfield  {journal} {\bibinfo  {journal} {Phys. Rev. A}\ }\textbf {\bibinfo {volume} {109}},\ \bibinfo {pages} {L040401} (\bibinfo {year} {2024})}\BibitemShut {NoStop}%
\bibitem [{\citenamefont {Tarabunga}\ and\ \citenamefont {Tirrito}(2025)}]{tarabunga2025magic}%
  \BibitemOpen
  \bibfield  {author} {\bibinfo {author} {\bibfnamefont {P.~S.}\ \bibnamefont {Tarabunga}}\ and\ \bibinfo {author} {\bibfnamefont {E.}~\bibnamefont {Tirrito}},\ }\bibfield  {title} {\bibinfo {title} {Magic transition in measurement-only circuits},\ }\href {https://doi.org/10.1038/s41534-025-01104-y} {\bibfield  {journal} {\bibinfo  {journal} {npj Quantum Inf.}\ }\textbf {\bibinfo {volume} {11}},\ \bibinfo {pages} {166} (\bibinfo {year} {2025})}\BibitemShut {NoStop}%
\bibitem [{\citenamefont {Szombathy}\ \emph {et~al.}(2025)\citenamefont {Szombathy}, \citenamefont {Valli}, \citenamefont {Moca}, \citenamefont {Farkas},\ and\ \citenamefont {Zar{\'a}nd}}]{szombathy2025asymptotically}%
  \BibitemOpen
  \bibfield  {author} {\bibinfo {author} {\bibfnamefont {D.}~\bibnamefont {Szombathy}}, \bibinfo {author} {\bibfnamefont {A.}~\bibnamefont {Valli}}, \bibinfo {author} {\bibfnamefont {C.~P.}\ \bibnamefont {Moca}}, \bibinfo {author} {\bibfnamefont {L.}~\bibnamefont {Farkas}},\ and\ \bibinfo {author} {\bibfnamefont {G.}~\bibnamefont {Zar{\'a}nd}},\ }\bibfield  {title} {\bibinfo {title} {Asymptotically independent fluctuations of stabilizer {R}{\'e}nyi entropy and entanglement in random unitary circuits},\ }\href {https://doi.org/10.1103/jplh-zl35} {\bibfield  {journal} {\bibinfo  {journal} {Phys. Rev. Res.}\ }\textbf {\bibinfo {volume} {7}},\ \bibinfo {pages} {043072} (\bibinfo {year} {2025})}\BibitemShut {NoStop}%
\bibitem [{\citenamefont {Masot-Llima}\ and\ \citenamefont {Garcia-Saez}(2024)}]{masot2024stabilizer}%
  \BibitemOpen
  \bibfield  {author} {\bibinfo {author} {\bibfnamefont {S.}~\bibnamefont {Masot-Llima}}\ and\ \bibinfo {author} {\bibfnamefont {A.}~\bibnamefont {Garcia-Saez}},\ }\bibfield  {title} {\bibinfo {title} {Stabilizer tensor networks: Universal quantum simulator on a basis of stabilizer states},\ }\href {https://doi.org/10.1103/PhysRevLett.133.230601} {\bibfield  {journal} {\bibinfo  {journal} {Phys. Rev. Lett.}\ }\textbf {\bibinfo {volume} {133}},\ \bibinfo {pages} {230601} (\bibinfo {year} {2024})}\BibitemShut {NoStop}%
\bibitem [{\citenamefont {Mittal}\ and\ \citenamefont {Huang}(2026)}]{mittal2026quantum}%
  \BibitemOpen
  \bibfield  {author} {\bibinfo {author} {\bibfnamefont {V.}~\bibnamefont {Mittal}}\ and\ \bibinfo {author} {\bibfnamefont {Y.-P.}\ \bibnamefont {Huang}},\ }\bibfield  {title} {\bibinfo {title} {Quantum magic in discrete-time quantum walk},\ }\href {https://doi.org/10.1103/7rwg-lhpv} {\bibfield  {journal} {\bibinfo  {journal} {Phys. Rev. Res.}\ }\textbf {\bibinfo {volume} {8}},\ \bibinfo {pages} {013124} (\bibinfo {year} {2026})}\BibitemShut {NoStop}%
\bibitem [{\citenamefont {Swingle}\ \emph {et~al.}(2016)\citenamefont {Swingle}, \citenamefont {Bentsen}, \citenamefont {Schleier-Smith},\ and\ \citenamefont {Hayden}}]{swingle2016measuring}%
  \BibitemOpen
  \bibfield  {author} {\bibinfo {author} {\bibfnamefont {B.}~\bibnamefont {Swingle}}, \bibinfo {author} {\bibfnamefont {G.}~\bibnamefont {Bentsen}}, \bibinfo {author} {\bibfnamefont {M.}~\bibnamefont {Schleier-Smith}},\ and\ \bibinfo {author} {\bibfnamefont {P.}~\bibnamefont {Hayden}},\ }\bibfield  {title} {\bibinfo {title} {Measuring the scrambling of quantum information},\ }\href {https://doi.org/10.1103/PhysRevA.94.040302} {\bibfield  {journal} {\bibinfo  {journal} {Phys. Rev. A}\ }\textbf {\bibinfo {volume} {94}},\ \bibinfo {pages} {040302} (\bibinfo {year} {2016})}\BibitemShut {NoStop}%
\bibitem [{\citenamefont {Landsman}\ \emph {et~al.}(2019)\citenamefont {Landsman}, \citenamefont {Figgatt}, \citenamefont {Schuster}, \citenamefont {Linke}, \citenamefont {Yoshida}, \citenamefont {Yao},\ and\ \citenamefont {Monroe}}]{landsman2019verified}%
  \BibitemOpen
  \bibfield  {author} {\bibinfo {author} {\bibfnamefont {K.~A.}\ \bibnamefont {Landsman}}, \bibinfo {author} {\bibfnamefont {C.}~\bibnamefont {Figgatt}}, \bibinfo {author} {\bibfnamefont {T.}~\bibnamefont {Schuster}}, \bibinfo {author} {\bibfnamefont {N.~M.}\ \bibnamefont {Linke}}, \bibinfo {author} {\bibfnamefont {B.}~\bibnamefont {Yoshida}}, \bibinfo {author} {\bibfnamefont {N.~Y.}\ \bibnamefont {Yao}},\ and\ \bibinfo {author} {\bibfnamefont {C.}~\bibnamefont {Monroe}},\ }\bibfield  {title} {\bibinfo {title} {Verified quantum information scrambling},\ }\href {https://doi.org/10.1038/s41586-019-0952-6} {\bibfield  {journal} {\bibinfo  {journal} {Nature}\ }\textbf {\bibinfo {volume} {567}},\ \bibinfo {pages} {61} (\bibinfo {year} {2019})}\BibitemShut {NoStop}%
\bibitem [{\citenamefont {Bertini}\ and\ \citenamefont {Piroli}(2020)}]{bertini2020scrambling}%
  \BibitemOpen
  \bibfield  {author} {\bibinfo {author} {\bibfnamefont {B.}~\bibnamefont {Bertini}}\ and\ \bibinfo {author} {\bibfnamefont {L.}~\bibnamefont {Piroli}},\ }\bibfield  {title} {\bibinfo {title} {Scrambling in random unitary circuits: Exact results},\ }\href {https://doi.org/10.1103/PhysRevB.102.064305} {\bibfield  {journal} {\bibinfo  {journal} {Phys. Rev. B}\ }\textbf {\bibinfo {volume} {102}},\ \bibinfo {pages} {064305} (\bibinfo {year} {2020})}\BibitemShut {NoStop}%
\bibitem [{\citenamefont {Mi}\ \emph {et~al.}(2021)\citenamefont {Mi}, \citenamefont {Roushan}, \citenamefont {Quintana}, \citenamefont {Mandra}, \citenamefont {Marshall}, \citenamefont {Neill}, \citenamefont {Arute}, \citenamefont {Arya}, \citenamefont {Atalaya}, \citenamefont {Babbush} \emph {et~al.}}]{mi2021information}%
  \BibitemOpen
  \bibfield  {author} {\bibinfo {author} {\bibfnamefont {X.}~\bibnamefont {Mi}}, \bibinfo {author} {\bibfnamefont {P.}~\bibnamefont {Roushan}}, \bibinfo {author} {\bibfnamefont {C.}~\bibnamefont {Quintana}}, \bibinfo {author} {\bibfnamefont {S.}~\bibnamefont {Mandra}}, \bibinfo {author} {\bibfnamefont {J.}~\bibnamefont {Marshall}}, \bibinfo {author} {\bibfnamefont {C.}~\bibnamefont {Neill}}, \bibinfo {author} {\bibfnamefont {F.}~\bibnamefont {Arute}}, \bibinfo {author} {\bibfnamefont {K.}~\bibnamefont {Arya}}, \bibinfo {author} {\bibfnamefont {J.}~\bibnamefont {Atalaya}}, \bibinfo {author} {\bibfnamefont {R.}~\bibnamefont {Babbush}}, \emph {et~al.},\ }\bibfield  {title} {\bibinfo {title} {Information scrambling in quantum circuits},\ }\href {https://doi.org/10.1126/science.abg5029} {\bibfield  {journal} {\bibinfo  {journal} {Science}\ }\textbf {\bibinfo {volume} {374}},\ \bibinfo {pages} {1479} (\bibinfo {year} {2021})}\BibitemShut {NoStop}%
\bibitem [{\citenamefont {Zhou}\ and\ \citenamefont {Luitz}(2017)}]{zhou2017operator}%
  \BibitemOpen
  \bibfield  {author} {\bibinfo {author} {\bibfnamefont {T.}~\bibnamefont {Zhou}}\ and\ \bibinfo {author} {\bibfnamefont {D.~J.}\ \bibnamefont {Luitz}},\ }\bibfield  {title} {\bibinfo {title} {Operator entanglement entropy of the time evolution operator in chaotic systems},\ }\href {https://doi.org/10.1103/PhysRevB.95.094206} {\bibfield  {journal} {\bibinfo  {journal} {Phys. Rev. B}\ }\textbf {\bibinfo {volume} {95}},\ \bibinfo {pages} {094206} (\bibinfo {year} {2017})}\BibitemShut {NoStop}%
\bibitem [{\citenamefont {Knap}(2018)}]{knap2018entanglement}%
  \BibitemOpen
  \bibfield  {author} {\bibinfo {author} {\bibfnamefont {M.}~\bibnamefont {Knap}},\ }\bibfield  {title} {\bibinfo {title} {Entanglement production and information scrambling in a noisy spin system},\ }\href {https://doi.org/10.1103/PhysRevB.98.184416} {\bibfield  {journal} {\bibinfo  {journal} {Phys. Rev. B}\ }\textbf {\bibinfo {volume} {98}},\ \bibinfo {pages} {184416} (\bibinfo {year} {2018})}\BibitemShut {NoStop}%
\bibitem [{\citenamefont {Ahmadi}\ and\ \citenamefont {Greplova}(2024)}]{ahmadi2024quantifying}%
  \BibitemOpen
  \bibfield  {author} {\bibinfo {author} {\bibfnamefont {A.}~\bibnamefont {Ahmadi}}\ and\ \bibinfo {author} {\bibfnamefont {E.}~\bibnamefont {Greplova}},\ }\bibfield  {title} {\bibinfo {title} {Quantifying non-stabilizerness via information scrambling},\ }\href {https://doi.org/10.21468/SciPostPhys.16.2.043} {\bibfield  {journal} {\bibinfo  {journal} {SciPost Phys.}\ }\textbf {\bibinfo {volume} {16}},\ \bibinfo {pages} {043} (\bibinfo {year} {2024})}\BibitemShut {NoStop}%
\bibitem [{\citenamefont {Choi}\ \emph {et~al.}(2020)\citenamefont {Choi}, \citenamefont {Bao}, \citenamefont {Qi},\ and\ \citenamefont {Altman}}]{choi2020quantum}%
  \BibitemOpen
  \bibfield  {author} {\bibinfo {author} {\bibfnamefont {S.}~\bibnamefont {Choi}}, \bibinfo {author} {\bibfnamefont {Y.}~\bibnamefont {Bao}}, \bibinfo {author} {\bibfnamefont {X.-L.}\ \bibnamefont {Qi}},\ and\ \bibinfo {author} {\bibfnamefont {E.}~\bibnamefont {Altman}},\ }\bibfield  {title} {\bibinfo {title} {Quantum error correction in scrambling dynamics and measurement-induced phase transition},\ }\href {https://doi.org/10.1103/PhysRevLett.125.030505} {\bibfield  {journal} {\bibinfo  {journal} {Phys. Rev. Lett.}\ }\textbf {\bibinfo {volume} {125}},\ \bibinfo {pages} {030505} (\bibinfo {year} {2020})}\BibitemShut {NoStop}%
\bibitem [{\citenamefont {Shen}\ \emph {et~al.}(2020)\citenamefont {Shen}, \citenamefont {Zhang}, \citenamefont {You},\ and\ \citenamefont {Zhai}}]{shen2020information}%
  \BibitemOpen
  \bibfield  {author} {\bibinfo {author} {\bibfnamefont {H.}~\bibnamefont {Shen}}, \bibinfo {author} {\bibfnamefont {P.}~\bibnamefont {Zhang}}, \bibinfo {author} {\bibfnamefont {Y.-Z.}\ \bibnamefont {You}},\ and\ \bibinfo {author} {\bibfnamefont {H.}~\bibnamefont {Zhai}},\ }\bibfield  {title} {\bibinfo {title} {Information scrambling in quantum neural networks},\ }\href {https://doi.org/10.1103/PhysRevLett.124.200504} {\bibfield  {journal} {\bibinfo  {journal} {Phys. Rev. Lett.}\ }\textbf {\bibinfo {volume} {124}},\ \bibinfo {pages} {200504} (\bibinfo {year} {2020})}\BibitemShut {NoStop}%
\bibitem [{\citenamefont {Garcia}\ \emph {et~al.}(2022)\citenamefont {Garcia}, \citenamefont {Bu},\ and\ \citenamefont {Jaffe}}]{garcia2022quantifying}%
  \BibitemOpen
  \bibfield  {author} {\bibinfo {author} {\bibfnamefont {R.~J.}\ \bibnamefont {Garcia}}, \bibinfo {author} {\bibfnamefont {K.}~\bibnamefont {Bu}},\ and\ \bibinfo {author} {\bibfnamefont {A.}~\bibnamefont {Jaffe}},\ }\bibfield  {title} {\bibinfo {title} {Quantifying scrambling in quantum neural networks},\ }\href {https://doi.org/10.1007/JHEP03(2022)027} {\bibfield  {journal} {\bibinfo  {journal} {J. High Energy Phys.}\ }\textbf {\bibinfo {volume} {2022}}\bibinfo  {number} { (3)},\ \bibinfo {pages} {1}}\BibitemShut {NoStop}%
\bibitem [{\citenamefont {Wu}\ \emph {et~al.}(2021)\citenamefont {Wu}, \citenamefont {Zhang},\ and\ \citenamefont {Zhai}}]{wu2021scrambling}%
  \BibitemOpen
\bibfield  {number} {  }\bibfield  {author} {\bibinfo {author} {\bibfnamefont {Y.}~\bibnamefont {Wu}}, \bibinfo {author} {\bibfnamefont {P.}~\bibnamefont {Zhang}},\ and\ \bibinfo {author} {\bibfnamefont {H.}~\bibnamefont {Zhai}},\ }\bibfield  {title} {\bibinfo {title} {Scrambling ability of quantum neural network architectures},\ }\href {https://doi.org/10.1103/PhysRevResearch.3.L032057} {\bibfield  {journal} {\bibinfo  {journal} {Phys. Rev. Res.}\ }\textbf {\bibinfo {volume} {3}},\ \bibinfo {pages} {L032057} (\bibinfo {year} {2021})}\BibitemShut {NoStop}%
\bibitem [{\citenamefont {Mohseni}\ \emph {et~al.}(2024)\citenamefont {Mohseni}, \citenamefont {Shi}, \citenamefont {Byrnes},\ and\ \citenamefont {Hartmann}}]{mohseni2024deep}%
  \BibitemOpen
  \bibfield  {author} {\bibinfo {author} {\bibfnamefont {N.}~\bibnamefont {Mohseni}}, \bibinfo {author} {\bibfnamefont {J.}~\bibnamefont {Shi}}, \bibinfo {author} {\bibfnamefont {T.}~\bibnamefont {Byrnes}},\ and\ \bibinfo {author} {\bibfnamefont {M.~J.}\ \bibnamefont {Hartmann}},\ }\bibfield  {title} {\bibinfo {title} {Deep learning of many-body observables and quantum information scrambling},\ }\href {https://doi.org/10.22331/q-2024-07-18-1417} {\bibfield  {journal} {\bibinfo  {journal} {Quantum}\ }\textbf {\bibinfo {volume} {8}},\ \bibinfo {pages} {1417} (\bibinfo {year} {2024})}\BibitemShut {NoStop}%
\bibitem [{\citenamefont {Oliviero}\ \emph {et~al.}(2024)\citenamefont {Oliviero}, \citenamefont {Leone}, \citenamefont {Lloyd},\ and\ \citenamefont {Hamma}}]{oliviero2024unscrambling}%
  \BibitemOpen
  \bibfield  {author} {\bibinfo {author} {\bibfnamefont {S.~F.}\ \bibnamefont {Oliviero}}, \bibinfo {author} {\bibfnamefont {L.}~\bibnamefont {Leone}}, \bibinfo {author} {\bibfnamefont {S.}~\bibnamefont {Lloyd}},\ and\ \bibinfo {author} {\bibfnamefont {A.}~\bibnamefont {Hamma}},\ }\bibfield  {title} {\bibinfo {title} {Unscrambling quantum information with {C}lifford decoders},\ }\href {https://doi.org/10.1103/PhysRevLett.132.080402} {\bibfield  {journal} {\bibinfo  {journal} {Phys. Rev. Lett.}\ }\textbf {\bibinfo {volume} {132}},\ \bibinfo {pages} {080402} (\bibinfo {year} {2024})}\BibitemShut {NoStop}%
\bibitem [{\citenamefont {Garcia}\ \emph {et~al.}(2023)\citenamefont {Garcia}, \citenamefont {Bu},\ and\ \citenamefont {Jaffe}}]{garcia2023resource}%
  \BibitemOpen
  \bibfield  {author} {\bibinfo {author} {\bibfnamefont {R.~J.}\ \bibnamefont {Garcia}}, \bibinfo {author} {\bibfnamefont {K.}~\bibnamefont {Bu}},\ and\ \bibinfo {author} {\bibfnamefont {A.}~\bibnamefont {Jaffe}},\ }\bibfield  {title} {\bibinfo {title} {Resource theory of quantum scrambling},\ }\href {https://doi.org/10.1073/pnas.2217031120} {\bibfield  {journal} {\bibinfo  {journal} {PNAS}\ }\textbf {\bibinfo {volume} {120}},\ \bibinfo {pages} {e2217031120} (\bibinfo {year} {2023})}\BibitemShut {NoStop}%
\bibitem [{\citenamefont {Bu}\ \emph {et~al.}(2024)\citenamefont {Bu}, \citenamefont {Garcia}, \citenamefont {Jaffe}, \citenamefont {Koh},\ and\ \citenamefont {Li}}]{bu2024complexity}%
  \BibitemOpen
  \bibfield  {author} {\bibinfo {author} {\bibfnamefont {K.}~\bibnamefont {Bu}}, \bibinfo {author} {\bibfnamefont {R.~J.}\ \bibnamefont {Garcia}}, \bibinfo {author} {\bibfnamefont {A.}~\bibnamefont {Jaffe}}, \bibinfo {author} {\bibfnamefont {D.~E.}\ \bibnamefont {Koh}},\ and\ \bibinfo {author} {\bibfnamefont {L.}~\bibnamefont {Li}},\ }\bibfield  {title} {\bibinfo {title} {Complexity of quantum circuits via sensitivity, magic, and coherence},\ }\href {https://doi.org/10.1007/s00220-024-05030-6} {\bibfield  {journal} {\bibinfo  {journal} {Commun. Math. Phys.}\ }\textbf {\bibinfo {volume} {405}},\ \bibinfo {pages} {161} (\bibinfo {year} {2024})}\BibitemShut {NoStop}%
\bibitem [{\citenamefont {Rall}\ \emph {et~al.}(2019)\citenamefont {Rall}, \citenamefont {Liang}, \citenamefont {Cook},\ and\ \citenamefont {Kretschmer}}]{rall2019simulation}%
  \BibitemOpen
  \bibfield  {author} {\bibinfo {author} {\bibfnamefont {P.}~\bibnamefont {Rall}}, \bibinfo {author} {\bibfnamefont {D.}~\bibnamefont {Liang}}, \bibinfo {author} {\bibfnamefont {J.}~\bibnamefont {Cook}},\ and\ \bibinfo {author} {\bibfnamefont {W.}~\bibnamefont {Kretschmer}},\ }\bibfield  {title} {\bibinfo {title} {Simulation of qubit quantum circuits via {P}auli propagation},\ }\href {https://doi.org/10.1103/PhysRevA.99.062337} {\bibfield  {journal} {\bibinfo  {journal} {Phys. Rev. A}\ }\textbf {\bibinfo {volume} {99}},\ \bibinfo {pages} {062337} (\bibinfo {year} {2019})}\BibitemShut {NoStop}%
\bibitem [{\citenamefont {Rudolph}\ \emph {et~al.}(2025)\citenamefont {Rudolph}, \citenamefont {Jones}, \citenamefont {Teng}, \citenamefont {Angrisani},\ and\ \citenamefont {Holmes}}]{rudolph2025pauli}%
  \BibitemOpen
  \bibfield  {author} {\bibinfo {author} {\bibfnamefont {M.~S.}\ \bibnamefont {Rudolph}}, \bibinfo {author} {\bibfnamefont {T.}~\bibnamefont {Jones}}, \bibinfo {author} {\bibfnamefont {Y.}~\bibnamefont {Teng}}, \bibinfo {author} {\bibfnamefont {A.}~\bibnamefont {Angrisani}},\ and\ \bibinfo {author} {\bibfnamefont {Z.}~\bibnamefont {Holmes}},\ }\bibfield  {title} {\bibinfo {title} {Pauli propagation: A computational framework for simulating quantum systems},\ }\href {https://doi.org/10.48550/arXiv.2505.21606} {\bibfield  {journal} {\bibinfo  {journal} {arXiv preprint arXiv:2505.21606}\ } (\bibinfo {year} {2025})}\BibitemShut {NoStop}%
\bibitem [{\citenamefont {Angrisani}\ \emph {et~al.}(2025)\citenamefont {Angrisani}, \citenamefont {Schmidhuber}, \citenamefont {Rudolph}, \citenamefont {Cerezo}, \citenamefont {Holmes},\ and\ \citenamefont {Huang}}]{angrisani2025classically}%
  \BibitemOpen
  \bibfield  {author} {\bibinfo {author} {\bibfnamefont {A.}~\bibnamefont {Angrisani}}, \bibinfo {author} {\bibfnamefont {A.}~\bibnamefont {Schmidhuber}}, \bibinfo {author} {\bibfnamefont {M.~S.}\ \bibnamefont {Rudolph}}, \bibinfo {author} {\bibfnamefont {M.}~\bibnamefont {Cerezo}}, \bibinfo {author} {\bibfnamefont {Z.}~\bibnamefont {Holmes}},\ and\ \bibinfo {author} {\bibfnamefont {H.-Y.}\ \bibnamefont {Huang}},\ }\bibfield  {title} {\bibinfo {title} {Classically estimating observables of noiseless quantum circuits},\ }\href {https://doi.org/10.1103/lh6x-7rc3} {\bibfield  {journal} {\bibinfo  {journal} {Phys. Rev. Lett.}\ }\textbf {\bibinfo {volume} {135}},\ \bibinfo {pages} {170602} (\bibinfo {year} {2025})}\BibitemShut {NoStop}%
\bibitem [{\citenamefont {Shao}\ \emph {et~al.}(2024)\citenamefont {Shao}, \citenamefont {Wei}, \citenamefont {Cheng},\ and\ \citenamefont {Liu}}]{shao2024simulating}%
  \BibitemOpen
  \bibfield  {author} {\bibinfo {author} {\bibfnamefont {Y.}~\bibnamefont {Shao}}, \bibinfo {author} {\bibfnamefont {F.}~\bibnamefont {Wei}}, \bibinfo {author} {\bibfnamefont {S.}~\bibnamefont {Cheng}},\ and\ \bibinfo {author} {\bibfnamefont {Z.}~\bibnamefont {Liu}},\ }\bibfield  {title} {\bibinfo {title} {Simulating noisy variational quantum algorithms: A polynomial approach},\ }\href {https://doi.org/10.1103/PhysRevLett.133.120603} {\bibfield  {journal} {\bibinfo  {journal} {Phys. Rev. Lett.}\ }\textbf {\bibinfo {volume} {133}},\ \bibinfo {pages} {120603} (\bibinfo {year} {2024})}\BibitemShut {NoStop}%
\bibitem [{\citenamefont {Fontana}\ \emph {et~al.}(2025)\citenamefont {Fontana}, \citenamefont {Rudolph}, \citenamefont {Duncan}, \citenamefont {Rungger},\ and\ \citenamefont {C{\^\i}rstoiu}}]{fontana2025classical}%
  \BibitemOpen
  \bibfield  {author} {\bibinfo {author} {\bibfnamefont {E.}~\bibnamefont {Fontana}}, \bibinfo {author} {\bibfnamefont {M.~S.}\ \bibnamefont {Rudolph}}, \bibinfo {author} {\bibfnamefont {R.}~\bibnamefont {Duncan}}, \bibinfo {author} {\bibfnamefont {I.}~\bibnamefont {Rungger}},\ and\ \bibinfo {author} {\bibfnamefont {C.}~\bibnamefont {C{\^\i}rstoiu}},\ }\bibfield  {title} {\bibinfo {title} {Classical simulations of noisy variational quantum circuits},\ }\href {https://doi.org/10.1038/s41534-024-00955-1} {\bibfield  {journal} {\bibinfo  {journal} {npj Quantum Inf.}\ }\textbf {\bibinfo {volume} {11}},\ \bibinfo {pages} {84} (\bibinfo {year} {2025})}\BibitemShut {NoStop}%
\bibitem [{\citenamefont {Roberts}\ \emph {et~al.}(2018)\citenamefont {Roberts}, \citenamefont {Stanford},\ and\ \citenamefont {Streicher}}]{roberts2018operator}%
  \BibitemOpen
  \bibfield  {author} {\bibinfo {author} {\bibfnamefont {D.~A.}\ \bibnamefont {Roberts}}, \bibinfo {author} {\bibfnamefont {D.}~\bibnamefont {Stanford}},\ and\ \bibinfo {author} {\bibfnamefont {A.}~\bibnamefont {Streicher}},\ }\bibfield  {title} {\bibinfo {title} {Operator growth in the {SYK} model},\ }\href {https://doi.org/10.1007/JHEP06(2018)122} {\bibfield  {journal} {\bibinfo  {journal} {J. High Energy Phys.}\ }\textbf {\bibinfo {volume} {2018}}\bibinfo  {number} { (6)},\ \bibinfo {pages} {1}}\BibitemShut {NoStop}%
\bibitem [{\citenamefont {Qi}\ \emph {et~al.}(2019)\citenamefont {Qi}, \citenamefont {Davis}, \citenamefont {Periwal},\ and\ \citenamefont {Schleier-Smith}}]{qi2019measuring}%
  \BibitemOpen
\bibfield  {number} {  }\bibfield  {author} {\bibinfo {author} {\bibfnamefont {X.-L.}\ \bibnamefont {Qi}}, \bibinfo {author} {\bibfnamefont {E.~J.}\ \bibnamefont {Davis}}, \bibinfo {author} {\bibfnamefont {A.}~\bibnamefont {Periwal}},\ and\ \bibinfo {author} {\bibfnamefont {M.}~\bibnamefont {Schleier-Smith}},\ }\bibfield  {title} {\bibinfo {title} {Measuring operator size growth in quantum quench experiments},\ }\href {https://doi.org/10.48550/arXiv.1906.00524} {\bibfield  {journal} {\bibinfo  {journal} {arXiv preprint arXiv:1906.00524}\ } (\bibinfo {year} {2019})}\BibitemShut {NoStop}%
\bibitem [{\citenamefont {Dowling}\ \emph {et~al.}(2025{\natexlab{b}})\citenamefont {Dowling}, \citenamefont {Kos},\ and\ \citenamefont {Turkeshi}}]{dowling2025magic}%
  \BibitemOpen
  \bibfield  {author} {\bibinfo {author} {\bibfnamefont {N.}~\bibnamefont {Dowling}}, \bibinfo {author} {\bibfnamefont {P.}~\bibnamefont {Kos}},\ and\ \bibinfo {author} {\bibfnamefont {X.}~\bibnamefont {Turkeshi}},\ }\bibfield  {title} {\bibinfo {title} {Magic resources of the {H}eisenberg picture},\ }\href {https://doi.org/10.1103/p7xt-s9nz} {\bibfield  {journal} {\bibinfo  {journal} {Phys. Rev. Lett.}\ }\textbf {\bibinfo {volume} {135}},\ \bibinfo {pages} {050401} (\bibinfo {year} {2025}{\natexlab{b}})}\BibitemShut {NoStop}%
\bibitem [{\citenamefont {Kaneyasu}\ and\ \citenamefont {Hasegawa}(2026)}]{availablity}%
  \BibitemOpen
  \bibfield  {author} {\bibinfo {author} {\bibfnamefont {M.}~\bibnamefont {Kaneyasu}}\ and\ \bibinfo {author} {\bibfnamefont {Y.}~\bibnamefont {Hasegawa}},\ }\href@noop {} {\bibinfo {title} {Scrambling trade-off}},\ \bibinfo {howpublished} {\url{https://github.com/chanmaoooo/scrambling_tradeoff}} (\bibinfo {year} {2026}),\ \bibinfo {note} {{G}it{H}ub repository}\BibitemShut {NoStop}%
\end{thebibliography}
\end{document}